\documentclass[twocolumn]{aastex702} 

\usepackage{float} 
\usepackage{amsmath}
\usepackage{epsf}
\usepackage{color}
\usepackage{graphicx}
\usepackage{color}
\usepackage{enumitem}
\usepackage{mathtools}
\usepackage[mathcal]{eucal}
 \let\mathscr\relax
\usepackage[scr]{rsfso}
\usepackage{ulem} 
\usepackage{cancel}
\usepackage{textcase}
\usepackage{subcaption}

\usepackage{tikz}
\usetikzlibrary{arrows,calc,positioning}

\tikzstyle{intt}=[draw,text centered,minimum size=6em,text width=3.25cm,text height=0.34cm]
\tikzstyle{intl}=[draw,text centered,minimum size=2em,text width=2.75cm,text height=0.34cm]
\tikzstyle{int}=[draw,minimum size=2.5em,text centered,text width=3.5cm]
\tikzstyle{intg}=[draw,minimum size=3em,text centered,text width=3.5cm]
\tikzstyle{sum}=[draw,shape=circle,inner sep=2pt,text centered,node distance=3.5cm]
\tikzstyle{summ}=[drawshape=circle,inner sep=4pt,text centered,node distance=3.cm]

\shorttitle{Relationship Between Bars, Star Formation Activity, and Host Galaxy Properties}
\shortauthors{Wise, Jogee, Guo et al.}

\def\gtsim{\lower.5ex\hbox{$\buildrel > \over\sim$}}
\def\gtrsim{\mathrel{\hbox{\rlap{\hbox{\lower4pt\hbox{$\sim$}}}\hbox{$>$}}}}
\def\lesssim{\mathrel{\hbox{\rlap{\hbox{\lower4pt\hbox{$\sim$}}}\hbox{$<$}}}}
\def\ltsim{\lower.5ex\hbox{$\buildrel < \over\sim$}}
\def\simgt{{\raise-.5ex\hbox{$\buildrel>\over\sim$}}\ } 
\def\simlt{{\raise-.5ex\hbox{$\buildrel<\over\sim$}}\ }

\def\sun{\mbox{$_\odot$}}

\def\e#1{\underline{#1}}  

\usepackage{graphicx}	
\usepackage{amsmath}	
\usepackage{amssymb}	

\begin{document}
\title{Exploring the Relationship Between Bars, Star Formation Activity, and Host Galaxy Properties from $\mathbf{z \sim 0}$ to $\mathbf{z \sim 2}$}

\author[0009-0009-8474-9361]{Eden Wise}
\affiliation{Department of Astronomy, The University of Texas at Austin, Austin, TX, USA}
\affiliation{Cosmic Frontier Center, The University of Texas at Austin, Austin, TX, USA}
\email{edenwise@utexas.edu}

\author[0000-0002-1590-0568]{Shardha Jogee}
\affiliation{Department of Astronomy, The University of Texas at Austin, Austin, TX, USA}
\affiliation{Cosmic Frontier Center, The University of Texas at Austin, Austin, TX, USA}
\email{sj@astro.as.utexas.edu}

\author[0000-0002-4162-6523]{Yuchen Guo}
\affiliation{Department of Astronomy, The University of Texas at Austin, Austin, TX, USA}
\affiliation{Cosmic Frontier Center, The University of Texas at Austin, Austin, TX, USA}
\email{kayguo98@utexas.edu}

\author[0000-0001-8519-1130]{Steven L. Finkelstein}
\affiliation{Department of Astronomy, The University of Texas at Austin, Austin, TX, USA}
\affiliation{Cosmic Frontier Center, The University of Texas at Austin, Austin, TX, USA}
\email{stevenf@astro.as.utexas.edu}

\author[0000-0002-9921-9218]{Micaela B. Bagley}
\affiliation{Department of Astronomy, The University of Texas at Austin, Austin, TX, USA}
\affiliation{Astrophysics Science Division, NASA Goddard Space Flight Center, 8800 Greenbelt Road, Greenbelt, MD 20771, USA}
\affiliation{Cosmic Frontier Center, The University of Texas at Austin, Austin, TX, USA}
\email{mbagley@utexas.edu}

\author[0000-0002-1416-8483]{Marc Huertas-Company}
\affiliation{Instituto de Astrofísica de Canarias, c/ Vía Láctea s/n, E-38205 La Laguna, Tenerife, Spain}
\affiliation{Departamento de Astrofísica, Universidad de La Laguna, E-38205 La Laguna, Tenerife, Spain}
\email{mhuertas@iac.es}

\author[0000-0001-9298-3523]{Kartheik G. Iyer}
\affiliation{Columbia Astrophysics Laboratory, Columbia University, 550 West 120th Street, New York, NY 10027, USA}
\affiliation{Center for Computational Astrophysics, Flatiron Institute, 162 Fifth Avenue, New York, NY 10010, USA}
\email{kgi2103@columbia.edu}

\author[0000-0001-8688-2443]{Elizabeth J.\ McGrath}
\affiliation{Department of Physics and Astronomy, Colby College, Waterville, ME 04901, USA}
\email{emcgrath@colby.edu}

\author[0000-0001-6917-4656]{Natalia C. Villanueva}
\affiliation{Department of Astronomy, The University of Texas at Austin, Austin, TX, USA}
\email{nataliavillanueva@utexas.edu}

\author[0000-0003-4244-8527]{Tommaso Zana}
\affiliation{Dipartimento di Fisica, Sapienza, Universit`a di Roma, Piazzale Aldo Moro 5, IT-00185 Roma, Italy}
\affiliation{INAF, Osservatorio Astronomico di Roma, Via di Frascati 33, IT-00078 Monte Porzio Catone, Italy}
\affiliation{INFN, Sezione di Roma I, Piazzale Aldo Moro 2, IT-00185 Roma, Italy}
\email{tommaso.zana@uniroma1.it}

\author[0000-0002-6610-2048]{Anton M. Koekemoer}
\affiliation{Space Telescope Science Institute, 3700 San Martin Drive, Baltimore, MD 21218, USA}
\email{koekemoer@stsci.edu}

\author[0000-0002-5564-9873]{Eric F. Bell}
\affil{Department of Astronomy, University of Michigan, Ann Arbor, MI, USA}
\email{ericbell@umich.edu}

\author[0009-0004-4168-3634]{Jean-Baptiste Billand}
\affil{Université Paris-Saclay, Université Paris Cité, CEA, CNRS, AIM, 91191 Gif-sur-Yvette, France}
\email{jean-baptiste.billand@cea.fr}

\author[0000-0001-8551-071X]{Yingjie Cheng}
\affiliation{Department of Astronomy, University of Washington, Seattle, WA 98195, USA}
\email{yingjiec@uw.edu}

\author[0000-0002-6219-5558, gname=Alexander, sname='de la Vega']{Alexander de la Vega}
\affiliation{Department of Physics and Astronomy, University of California, 900 University Ave, Riverside, CA 92521, USA}
\email{alexandd@ucr.edu}

\author[0000-0003-2676-8344]{Elena D’Onghia}
\affiliation{Department of Physics, University of Wisconsin-Madison, Madison, WI 53706, USA}
\affiliation{Department of Astronomy, University of Wisconsin-Madison, Madison, WI 53706, USA}
\email{edonghia@astro.wisc.edu}

\author[0000-0002-6851-9613]{Tobias G\'eron}
\affiliation{Dunlap Institute for Astronomy \& Astrophysics, University of Toronto, 50 St. George Street, Toronto, ON M5S 3H4, Canada}
\email{tobias.geron@utoronto.ca}

\author[0000-0002-7831-8751]{Mauro Giavalisco}
\affiliation{University of Massachusetts Amherst, 710 North Pleasant Street, Amherst, MA 01003-9305, USA}
\email{mauro@umass.edu}

\author[0000-0002-4884-6756]{Benne W. Holwerda}
\affil{Physics \& Astronomy Department, University of Louisville, 40292 KY, Louisville, USA}
\email{benne.holwerda@louisville.edu}

\author[0000-0003-1581-7825]{Ray A. Lucas}
\affiliation{Space Telescope Science Institute, 3700 San Martin Drive, Baltimore, MD 21218, USA}
\email{lucas@stsci.edu}

\author[0000-0001-9879-7780]{Fabio Pacucci}
\affiliation{Center for Astrophysics $\vert$ Harvard \& Smithsonian, Cambridge, MA 02138, USA}
\affiliation{Black Hole Initiative, Harvard University, Cambridge, MA 02138, USA}
\email{fabio.pacucci@cfa.harvard.edu}

\author[0009-0009-3123-4479]{Maxime Tarrasse}
\affil{Université Paris-Saclay, Université Paris Cité, CEA, CNRS, AIM, 91191 Gif-sur-Yvette, France}
\email{maxime.tarrasse@cea.fr}

\author[0000-0003-3466-035X]{{L. Y. Aaron} {Yung}}
\affiliation{Space Telescope Science Institute, 3700 San Martin Drive, Baltimore, MD 21218, USA}
\email{yung@stsci.edu}

\begin{abstract}

We present the most comprehensive study to date of the relationship between bars, star formation, and galaxy properties from $z \sim$~0 to $z \sim$~2. We use a mass-complete sample of 1,171 galaxies from the \textit{JWST} CEERS survey with $M_\star > 10^{10}\,M_\odot$ and repeat the analysis using COSMOS-Web data. Our results are: 1) At high redshift ($z \sim$~$1-2$) barred galaxies tend to have high sSFRs and low Sérsic indices ($n \leq 2$), while at low redshifts barred galaxies emerge with both low sSFR and higher $n$, suggestive of quiescent galaxies with bulges. 2) The fractional contribution of barred quiescent galaxies to the bar fraction rises steeply from $z \sim$~2 to $z \sim$~0, while that of barred actively star-forming galaxies falls. 3) The fraction of quiescent galaxies that are barred rises steeply over the last 10 Gyr. 4) Our empirical results show good agreement with the TNG50-1 simulations for bars with $a_{\mathrm{bar}}$ $>$ 1.5 kpc. Our results allow for the possibility that bar-driven secular evolution may lead to quiescence and/or that bars are more likely to persist and grow in gas-poor, quiescent galaxies. The steep rise in the quiescent bar fraction over 10 Gyr may represent an evolutionary sequence whereby gas-rich disks at high redshift first develop short, dynamically young bars and over time, repeated bar-driven gas inflows lead to central starbursts and declining gas fractions that strengthen the bar as the galaxy transitions toward quiescence.

\end{abstract}

\section{Introduction}\label{Sec: Introduction}
 
Stellar bars are considered to be powerful drivers of secular evolution in disk galaxies (e.g., \citealt{KormendyKennicutt2004} and references therein; \citealt{Athanassoula2003,Jogee2005}). Observations in the local Universe reveal that the majority of present-day spiral galaxies--including the
Milky Way--host stellar bars (e.g., \citealt{BlitzSpergel1991,Eskridge2000, MarinovaJogee2007, MenendezDelmestre2007,Geron2021}). These non-axisymmetric structures, dynamically defined by a family of periodic stellar orbits, exert gravitational torques and shocks that efficiently channel gas from the outer disk into the circumnuclear (CN) or inner kpc region, raising CN gas densities and triggering intense star formation (SF) (e.g., \citealt{Athanassoula1992b, KormendyKennicutt2004, Sakamoto1999, Jogee2005}). Over time, this process contributes to the buildup of central compact stellar disks known as disky bulges or pseudobulges (e.g., \citealt{Kormendy1993, KormendyKennicutt2004, Athanassoula2005, Jogee2005}). Bars are not expected to directly fuel AGN in present-day barred galaxies as bar-driven gas inflows tend to stall in the inner few hundred pc near the Inner Lindblad Resonances (ILRs) and additional transport mechanisms are needed to drive the gas down to the central black hole (\citealt{Buta1996}; review by \citealt{Jogee2006} and references therein). However, some studies do find that AGN host galaxies are marginally more likely to host a bar (e.g., \citealt{Galloway2015, Garland2023}) and that barred active galaxies experience excess nuclear activity in comparison to unbarred active galaxies (e.g., \citealt{Alonso2013,Marels2025}). The relationship between AGN feedback and the susceptibility of disks to bar formation is a topic of ongoing study (e.g., \citealt{Lokas2022,Rosas2025}).

Bars have been posited to play a role in driving quiescence in galaxies (e.g., \citealt{Masters2010b, Masters2012, Cheung2013,Geron2021}). One possible pathway for this to happen is that over time, the repeated effects of bar-driven gas inflow and accelerated SF activity in CN regions can help make a galaxy quiescent by redistributing the gas from the outer disk into the CN region and accelerating the conversion of this gas into stars. Once the conditions for SF are no longer met in the CN region (e.g., due to lower gas densities or stellar/AGN feedback), the region can transition into a CN post-starburst phase. As long as the barred galaxy is accreting gas, the aforementioned cycle can repeat, with the galaxy moving from a bar-driven gas inflow phase to a CN starburst phase and subsequently to a CN post-starburst phase. We discuss this scenario in more detail in Section~\ref{sec:scen1} where the schematic figure shows the different phases in action. At low redshifts, once the barred galaxy stops accreting gas, the galaxy can stay in this relatively quiescent post-starburst phase.
 
However, mechanisms other than bars may also lead to such transformations. For example, the effects of tidal interactions can lead to large gas inflows, accelerated SF, bulge-building, and eventually quiescence (e.g., \citealt{Barnes1992} and references therein; \citealt{Gnedin2003}). The history of gas accretion over time and environmental effects in galaxy clusters, such as strangulation, starvation, and ram pressure stripping, can also contribute to inducing quiescence (e.g., \citealt{Gunn1972, Larson1980, Koopmann2004,Crowl2005,Singh2019,Gentile2025}).

The relationship between barred galaxies, SF activity, the onset of quiescence, and host galaxy properties has been studied at $z\sim0$ (e.g., \citealt{Jogee2005,Masters2010b, Masters2010a, Masters2012, Cheung2013, Ellison2011, Khoperskov2018, Lin2020, Fraser2020, Newnham2020, George2020, George2021}), with some studies extending out to intermediate redshifts, including \citet{Cameron2010} out to $z\sim0.6$ and \citet{Melvin2014} out to $z\sim1$, but little is known at higher redshifts. 

\citet{Ellison2011} report a modest enhancement in the SFRs of barred galaxies compared to unbarred galaxies, but the enhancement is only for galaxies with total stellar mass $M_\star > 10^{10}\ M_\odot$ and is not seen in lower mass systems.  
At $z\sim0$, \citet{Masters2010b} report a large optical bar fraction in red spirals (70\% $\pm$ 5\% versus 27\% $\pm$ 5\% in blue spirals) and suggest that the cessation of SF and bar instabilities in spirals are strongly correlated. \citet{Cheung2013} report that, at $z\sim0$, the likelihood of a galaxy hosting a bar is anticorrelated with specific star formation rate (sSFR), regardless of stellar mass or bulge prominence. Among $z\sim0$ galaxies, \citet{Masters2010a} report that there is a clear increase in the bar fraction with redder (g-r) colors, decreased luminosity, and in galaxies with more prominent bulges, to the extent that over half of the red, bulge-dominated disk galaxies in their sample possess a bar. \citet{George2020} also report that the regions near the bar in a number of barred galaxies are devoid of recent SF as well as molecular and neutral hydrogen. In addition, \citet{Cameron2010} find that the bar fraction in low sSFR disks at intermediate masses is higher than in high sSFR disks at $0.2 < z < 0.6$, and \citet{Melvin2014} report a high bar fraction of $45\%\pm5\%$ in a sample of 98 red disk galaxies at $0.4 \le z \le 1.0$.

However, no such explorations have been conducted at higher redshifts. In this paper, we perform one of the first explorations of the relationship between bars, SF activity, and host galaxy properties from $z\sim0$ to $z\sim2$. Such studies are possible thanks to the advent of \textit{Hubble Space Telescope} (\textit{HST}) and \textit{James Webb Space Telescope} (\textit{JWST}; \citealt{Gardner2006, Gardner2023, Rigby2023}) data. Two decades of \textit{HST} studies of barred galaxies have extended our understanding of bars out to $z \sim 1$ (e.g., \citealt{Abraham1999,Elmegreen2004,Jogee2004,Sheth2008,Melvin2014,Simmons2014}). With its sensitive, high-resolution, infrared imaging \citep{Rieke2023}, \textit{JWST} has enabled rest-frame near-infrared observations of galaxies beyond $z \sim 1$ for the first time, allowing us to trace older stellar populations with reduced sensitivity to dust and recent SF (e.g., \citealt{Frogel1996,Suess2022}). Early \textit{JWST} studies have extended the redshift frontier for barred galaxies out to $z \sim 4$ (e.g., \citealt{Guo2025,LeConte2024,Geron2025,HuertasCompany2025, Salcedo2025, LeConte2026}). In particular, \citet{Guo2025} provide compelling evidence for the early emergence of stellar bars, at least as early as $z \sim 4$, and a rise in the observed bar fraction, average projected bar length, and projected bar strength from $z \sim 4$ to $z \sim 0.5$. These results highlight the early emergence of barred galaxies and the increasing importance of bar-driven secular evolution from $z \sim 4$ (when the Universe was only $\sim 11\%$ of its present age) to the present day.

While these \textit{JWST} studies have placed constraints on the abundance and emergence of bars, the role of bars in regulating SF activity and driving galaxy evolution at high redshift remains poorly understood. One must also consider that the formation, evolution, and destruction of stellar bars depend on the interplay between the stellar disk, dark matter halo, and gas within galaxies (e.g., \citealt{Athanassoula2003,Athanassoula2013,Sellwood2016}). Gas can both promote the growth of $m=2$ bar instabilities in dynamically cold disks (e.g., \citealt{Bournaud2002,Romano2008,Kraljic2012,Spinoso2017,Rosas2022}) and weaken or destroy bars through the buildup of central mass concentrations (e.g., \citealt{Shen2004,Bournaud2005,Athanassoula2005cmc,Debattista2006}), while bars may form spontaneously (e.g., \citealt{Toomre1981,Ostriker1973,Athanassoula2003}; D'Onghia et al. 2026, in prep.) or be tidally induced through galaxy interactions (e.g., \citealt{Noguchi1987,Miwa1998,Bi2022,Rosas2024}). As gas fractions and environment evolve with cosmic time, both the SF activity and structural evolution of barred galaxies at earlier cosmic times may be significantly different from what is known at low redshifts.

This work is organized as follows: Section~\ref{Sec: observation and data} outlines the observational data and sample selection. In Sections~\ref{sec:vis class CEERS}--\ref{sec:bar id CEERS}, we describe the selection of moderately inclined disks and bars in CEERS. In Sections~\ref{sec:quiescent id} and~\ref{sec:starform id}, we discuss our methodology for identifying quiescent and actively star-forming disks. We then discuss the COSMOS-Web data we use as a secondary check on CEERS results in Section~\ref{sec:COSMOS check}. In Section~\ref{sec:host properties}, we explore the relationship between barred galaxies, sSFR, and Sérsic index. Section~\ref{sec:SF Q contribution} outlines the contribution of quiescent and actively star-forming galaxies to the total bar fractions in CEERS and COSMOS-Web. Section~\ref{sec:quiescent SF fbar} describes how the fraction of quiescent and actively star-forming disks that are barred evolves from $z \sim~2$  to $z \sim~0$, the effects of disk sizes on the quiescent bar fraction, and a comparison of our results to the IllustrisTNG TNG50-1 cosmological simulations \citep{Marinacci-etal-2018,Naiman-etal-2018,Nelson-etal-2018, Springel-etal-2018,Pillepich-etal-2018,Pillepich-etal-2019, Nelson-etal-2019,Nelson-etal-2019-Release}. In Section~\ref{sec:discussion}, we discuss the possible interpretations of our results on the relationship between bars and SF, particularly the increasing likelihood of barred galaxies to be quiescent systems with higher Sérsic indices, suggestive of central bulges.    

We assume the latest Planck flat $\Lambda$CDM cosmology with $H_0 = 67.36$ km$^{-1}$ Mpc$^{-1}$, $\Omega_m = 0.3153$, and $\Omega_\Lambda = 0.6847$ \citep{Planck-etal-20}. All magnitudes quoted are in the absolute bolometric (AB) system (\citealt{Oke-Gunn1983}).

\section{Observational data and sample selection} \label{Sec: observation and data}

\subsection{CEERS NIRCam Imaging}\label{Sec: CEERS Imaging}

This work leverages imaging from the Cosmic Evolution Early Release Science (CEERS) Survey \citep{Finkelstein2022, Finkelstein2025}. CEERS provides deep, high-resolution NIRCam imaging across 10 pointings in the Extended Groth Strip (EGS) \citep{Davis2007}, covering a total area of $\sim90$ arcmin$^2$. The observations span seven bands: F115W, F150W, F200W, F277W, F356W, F410M, and F444W, enabling rest-frame optical and near-infrared coverage out to  $z \sim 2$. Typical total exposure time for pixels observed in all three dithers was 2835s per band, with a 5$\sigma$ depth ranging from 28.8 to 29.7 \citep{Bagley2023, Finkelstein2025}. The CEERS images used in this work can be found on the CEERS website (\url{ceers.github.io/releases.html}) and on MAST via: \dataset[10.17909/z7p0-8481]{\doi{10.17909/z7p0-8481}}.

\citet{Bagley2023} performed the reduction of NIRCam images using version 1.8.5 of the \textit{JWST} Calibration Pipeline \citep{Bushouse2022}, which includes background subtraction, artifact mitigation, astrometric alignment, and mosaic creation. We use the photometric catalog described in \citet{Finkelstein2024}, which was produced using \texttt{Source Extractor} \citep{Bertin1996} v2.25.0 in two-image mode. The detection image was constructed from an inverse-variance weighted combination of the PSF-matched F277W and F356W images to optimize both depth and spatial resolution. Photometry was measured in all seven NIRCam bands on PSF-matched images. Photometric redshifts were estimated with \texttt{eazy} \citep{Brammer2008}, following the methodology in \citet{Finkelstein2023}. The sharp resolution of \textit{JWST} NIRCam (PSF FWHM of $\sim$ 0.08$"$ in F200W and 0.163$"$ in F444W) enables robust structural measurements of galaxies, including stellar bars, out to unprecedented redshifts.

\subsection{Sample Selection}\label{sec:sample selection}

We base our main analysis on the catalog of galaxies presented in \citet{Guo2025}, comprised of 1,770 galaxies with stellar mass \(M_\star \geq 10^{10}\,M_\odot\) and photometric redshifts in the range \(0.5 \leq z \leq 4.0\). This mass cut ensures completeness across the survey volume, while the redshift range traces the evolution of bars over \(\sim\!12\) Gyr of cosmic history. In this work, we use a subset of 1,171 galaxies at $z \sim 0.5$ to $z \sim 2$, tracing \(\sim\!10\) Gyr of cosmic history. Where available, \citet{Guo2025} supplemented photometric redshifts with spectroscopic redshifts in EGS (N. Hathi 2022, private communication) and spectroscopic redshifts derived from JWST NIRSpec observations (\citealt{Arrabal2023}; Arrabal Haro et al. 2025, in prep.). 

The identification of bars in this analysis is based on ellipse-fits performed by \citet{Guo2025}. Due to the resolution limits of the data and the nature of ellipse-fitting methods, the resulting bar sample mainly includes systems with projected semi-major axes of \(a_{\rm bar} > 1.5\) kpc (approximately 2 times the PSF in F200W). This scale corresponds to roughly two resolution elements along the semi-major length of the bar, which is the general requirement for robust identification of bars \citep{Erwin2018, Liang2024}. 

The stellar mass and SFR measurements used in this work are additionally drawn from the catalog presented in \citet{Guo2025}. Stellar masses and SFRs were provided by the CEERS team (Iyer et al. 2025, in prep.) and derived using the \textsc{Dense Basis} spectral energy distribution fitting code \citep{Iyer2019}, which accurately reconstructs non-parametric star formation histories via a fully Bayesian inference. The fits were run using v0.18 of \textsc{Dense Basis}, assuming a \citet{Calzetti2000} dust law and \citet{Chabrier2003} IMF, and using the same priors as described in \citet{Iyer2019}.

\section{Methodology}\label{sec:method}
In this methodology section, we first outline how we identify the sample of moderately inclined disks in CEERS  
(Sections~\ref{sec:vis class CEERS} and~\ref{sec:sersic CEERS}) from which bars will be identified. Next, we describe 
the identification and characterization of barred galaxies in CEERS (Section~\ref{sec:bar id CEERS}). We outline our methods to identify quiescent galaxies and star-forming galaxies in Section~\ref{sec:quiescent id} and Section~\ref{sec:starform id}, respectively. Finally, we describe the COSMOS-Web sample we utilize for cross-checks in Section~\ref{sec:COSMOS check}.

\subsection{Selection of Moderately Inclined Disk Galaxies in CEERS
}\label{sec:vis class CEERS}

In this work, we analyze a sample of moderately inclined disk galaxies presented in \citet{Guo2025}. We begin with this sample of disks in order to calculate bar fractions $f$$_{bar}$, defined as the fraction of disk galaxies that host stellar bars. We briefly summarize the methodology used by \citet{Guo2025} to identify moderately inclined disks in the following text; a complete description can be found in \citet{Guo2025}. 

Disk galaxies were first identified via visual classification. Three trained classifiers inspected multi-band NIRCam cutouts (F115W, F150W, F200W, F277W, F356W, F410M and F444W) for all galaxies. Each galaxy was classified as a disk if at least two out of three classifiers identified it as such based on the presence of an extended outer component. \citet{Guo2025} defines an extended outer disk component as a component with a slowly declining surface brightness profile which surrounds the central region of a galaxy.

In order to select moderately inclined disks suitable for bar identification, an axis ratio cut was applied using single-component Sérsic fits from \texttt{GALFIT} \citep{McGrath2026}. \citet{Guo2025} estimated inclinations using the relation $\cos(i) = b/a$, assuming a thin disk, and defined moderately inclined systems as those with $b/a > 0.42$ (i.e., $i \leq 65^\circ$) in at least one of the F277W, F356W, or F444W filters. This criterion yielded a final sample of 839 moderately inclined disk galaxies in \citet{Guo2025} (576 at $z \sim 0.5$ to $z \sim 2$), which form the main sample we use in our analysis.

\subsection{Distribution of Sérsic Indices Among Disk Galaxies in CEERS
}\label{sec:sersic CEERS}

In addition to visual classifications, we examine the structural properties of these galaxies to provide an independent and quantitative assessment of the disk population. We utilize Sérsic index measurements derived using \texttt{GALFIT} from \citet{McGrath2026}. The Sérsic index ($n$) is one of the most widely used single-parameter structural indicators of galaxy morphology \citep{Sersic1963}. 
Values of $n \lesssim 2.5$ typically correspond to shallow surface-brightness profiles characteristic of disk-dominated galaxies \citep{Ravindranath2004,Barden2005, Kartaltepe2023}. Higher values ($n \gtrsim 3$–$4$) can indicate a central compact steeply declining light profile surrounded by an extended outer envelope, as is seen in spheroids or galaxies hosting bulges (e.g., \citealt{Ravindranath2004,Bell2004,Trujillo2007}). Compact relic starbursts can also lead to high $n \gtrsim$ $4$ (e.g., \citealt{Hopkins2010}). Although the Sérsic index alone cannot fully capture the complexity of galaxy structure, it provides a valuable diagnostic for assessing the robustness of visual classifications across large samples and wide redshift ranges.

\begin{figure}[!htb]
         \centering
         \includegraphics[width=1\linewidth]{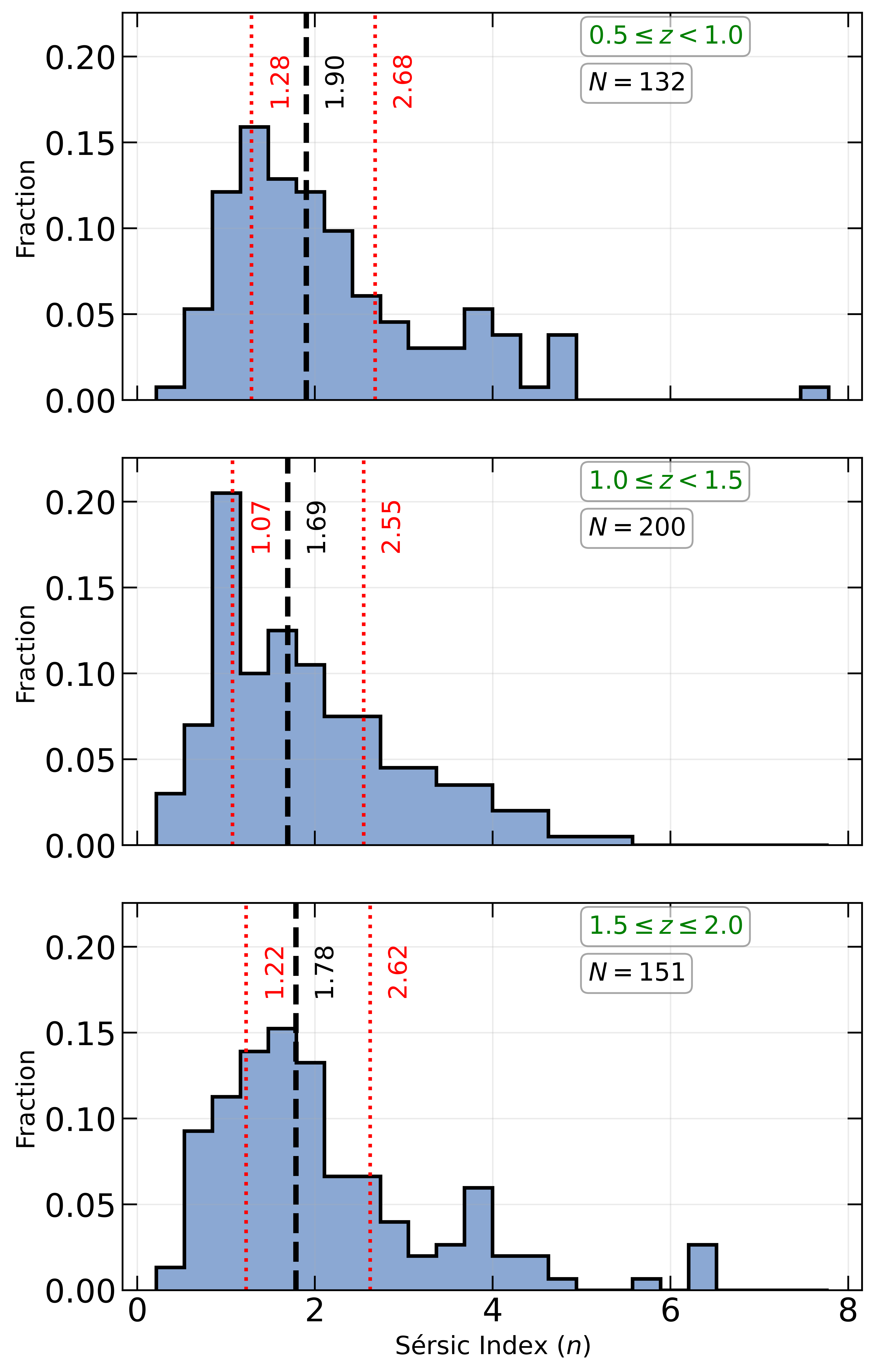}
         \caption{Distribution of Sérsic indices ($n$) for visually identified disk galaxies. The distributions peak at $n \le 2$, consistent with shallow light profiles expected of disk-dominated galaxies. The black dashed lines mark the median $n$ of the galaxy population in each redshift bin. The red dashed lines mark the 25th and 75th percentiles of the distribution in each bin, illustrating the interquartile range of the sample. The concentration of low-$n$ systems confirms that our visual classifications predominantly select structurally disk-dominated galaxies.}
    \label{fig:sersicn}
\end{figure} 

Figure~\ref{fig:sersicn} shows the distribution of Sérsic indices measured in F356W in each redshift bin (0.5 $\leq$$z$ $<$ 1.0, 1.0 $\leq$ $z$ $<$ 1.5, 1.5 $\leq$ $z$ $\leq$ 2.0) for galaxies visually identified as disks that have quality flags $<$ 2 in F356W \citep{McGrath2026}. The quality flags refer to the quality of the \texttt{GALFIT} fit; sources whose fits converged and whose model magnitude fell within 3 times the dispersion in the running median offset between \texttt{GALFIT} and SExtractor are considered reliable and assigned a flag of 0. A flag value of 1 corresponds to sources whose best-fit magnitudes did not fall within 3$\sigma$ of the running median offset, and a flag value of 2 corresponds to sources where one or more parameters reached a constraint limit during fitting. The distribution in each bin is strongly peaked at $n \le 2$, consistent with the shallow light profiles expected for disk-dominated systems, with median values (marked by the black dashed lines) of $n_{\mathrm{med}} = 1.9$, $1.69$, and $1.78$ in the three redshift bins, respectively. The red dashed lines mark the interquartile range ($n_{25}=1.28$, $1.07$, $1.22$, and $n_{75}=2.68$,  $2.55$, $2.62$), indicating that most of the disk sample lies well within the canonical disk regime ($n \lesssim 2.5$). Only a small fraction ($< 18\%$ in each bin) of visually identified disks extend to higher Sérsic indices ($n \gtrsim 3$), likely corresponding to galaxies hosting significant bulges or compact relic starbursts. Visual inspection of these sources indeed confirms the presence of prominent bulges and compact central stellar concentrations.

The consistency between the visually identified disk sample and the distribution of $n$ values demonstrates the reliability of the morphological classifications presented in \citet{Guo2025}.

\subsection{Identification and Characterization of Bars in CEERS
}\label{sec:bar id CEERS}

The ellipse-fit based bar classifications used in this work are those defined and applied in \citet{Guo2025}. We briefly summarize the bar identification methodology for ellipse-fit based classifications used by \citet{Guo2025} below; a full description can be found in \citet{Guo2025}.

We first emphasize that \citet{Guo2025} leveraged both F200W and F444W \textit{JWST} NIRCam images for bar identification. The shorter wavelength F200W filter offers sharper angular resolution (PSF FWHM $\sim$0.08$"$) but is more sensitive to younger stellar populations, tracing from a median rest-frame wavelength of 11428.57\,\AA\ in the lowest redshift bin to 7272.73\,\AA\ in the highest redshift bin. The longer wavelength F444W filter has a lower angular resolution, but is less affected by dust and recent SF and traces low and intermediate mass stars which make up the bulk of the stellar mass, tracing from a median rest-frame wavelength of 25371.43\,\AA\ in the lowest redshift bin to 16145.45\,\AA\ in the highest redshift bin. The use of both bands ensures that bars are robustly detected across a range of redshifts and stellar populations.

To identify bars \citet{Guo2025} performed ellipse fits following the methodology of \citet{Jogee2004} and
\citet{MarinovaJogee2007}, first ellipse fitting images \citep[e.g., ][]{Jedrzejewski1987, Wozniak1995, Jogee2002, Jogee2004, Elmegreen2004, MarinovaJogee2007}, and then applying quantitative criteria to identify bars. The criteria require the following: (a) over the bar region, the radial ellipticity profile rises smoothly to a significant peak ($e_{\rm max} \geq 0.25$) while the position angle stays relatively constant, and (b) as we transition from the bar to the region dominated by the outer disk, there is a drop in ellipticity ($>0.1$) and a significant change in position angle ($\gtrsim 10^\circ$); a more detailed discussion of these criteria can be found in Section 3.2.2 and Appendix A of \citet{Guo2025}. \citet{Guo2025} also visually inspected the fitted ellipses overlaid on the image of each galaxy to remove cases where bar signatures may have been mimicked by artifacts such as clumps, inclined rings, or spiral arms. 

In this paper, we use both the CEERS sample and the COSMOS-Web sample from \citet{HuertasCompany2025} (see Section~\ref{sec:COSMOS check}). A strong advantage of the CEERS sample is that the characterization of bars via ellipse fits in \citet{Guo2025} provides \textit{quantitative} bar properties (e.g., maximum ellipticity, semi-major axis). These quantitative properties allow us to address systematic effects tied to bar size and strength, perform tailored apples-to-apples comparisons with simulations, and study the evolution of barred galaxies over cosmic time. In contrast, such quantitative bar properties are not available in the COSMOS-Web study \citep{HuertasCompany2025} where bars are identified via neural networks trained on visual classifications.

\subsection{Identifying Quiescent Galaxies}\label{sec:quiescent id}

\begin{figure*}[!htb]
         \centering
         \includegraphics[width=1\textwidth]{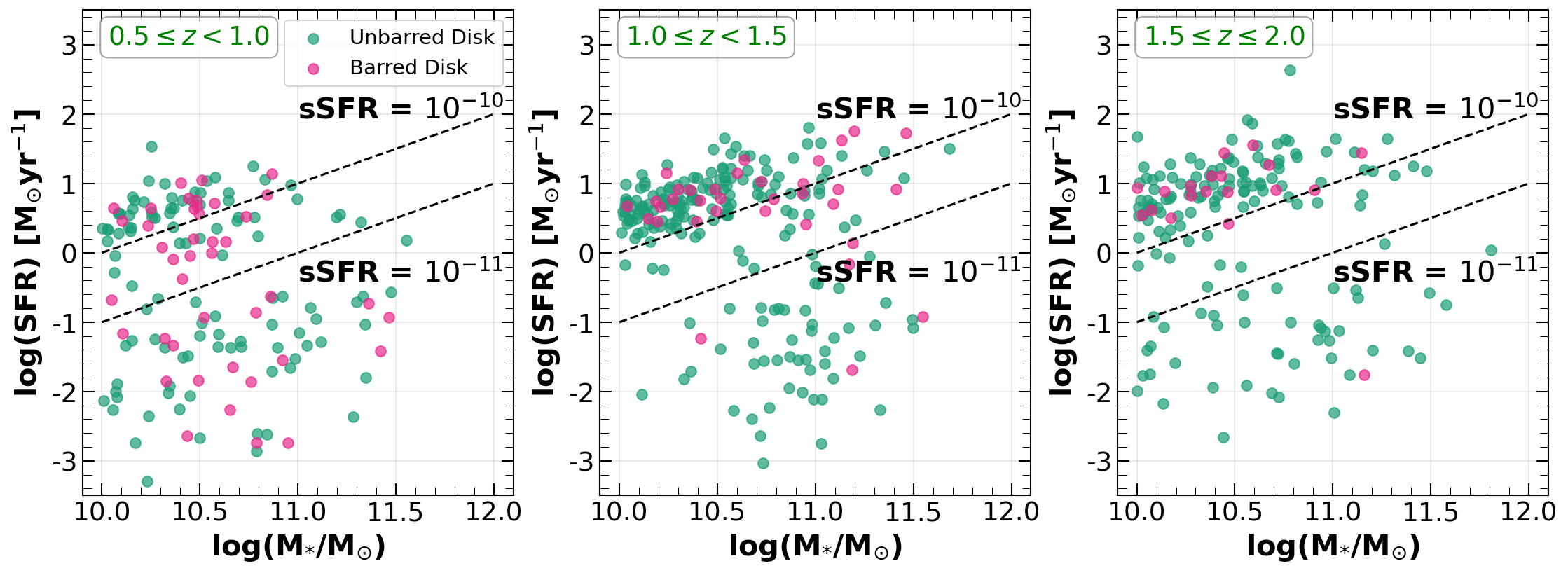}
         \caption{We define quiescent galaxies across all redshift bins as galaxies with sSFRs below 10$^{-11}$ yr$^{-1}$. This figure shows unbarred disk galaxies in green and barred disk galaxies in magenta, with the constant sSFR cut for quiescent galaxies shown as a dashed line labeled sSFR $=$ 10$^{-11}$ yr$^{-1}$. We also define actively star-forming galaxies as those with sSFRs above 10$^{-10}$ yr$^{-1}$.
}
\label{fig:sfr_mass}
\end{figure*}

 \begin{figure*}[!htb]
         \centering
         \includegraphics[width=1\textwidth]{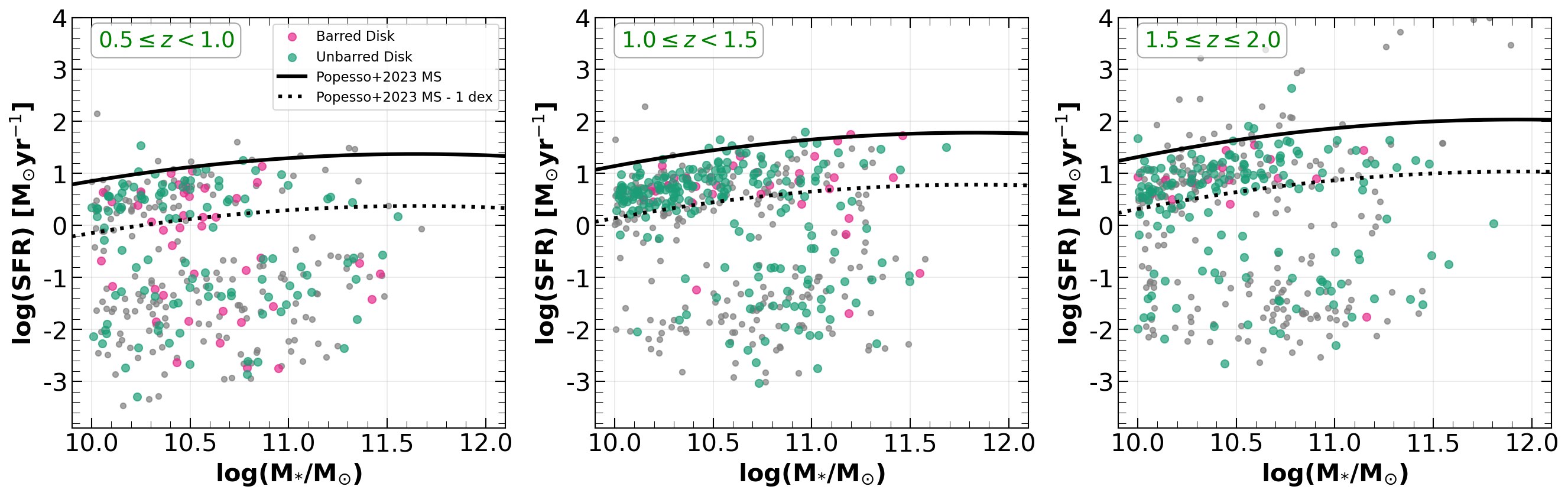}
         \caption{We define quiescent galaxies as below the main sequence -1 dex. This figure shows unbarred disk galaxies in green, barred disk galaxies in magenta, and galaxies above our stellar mass cut that do not fall within those definitions in grey (e.g., spheroids, highly inclined disks, etc.). The main sequence parameterization presented in Equation 10 and Table 2 of \citet{Popesso2023} is shown as the solid black line and this main sequence -1 dex is shown as the dashed black line.
}
\label{fig:CEERS_MS}
\end{figure*} 

\begin{figure*}[!htb]
         \centering
         \includegraphics[width=1\textwidth]{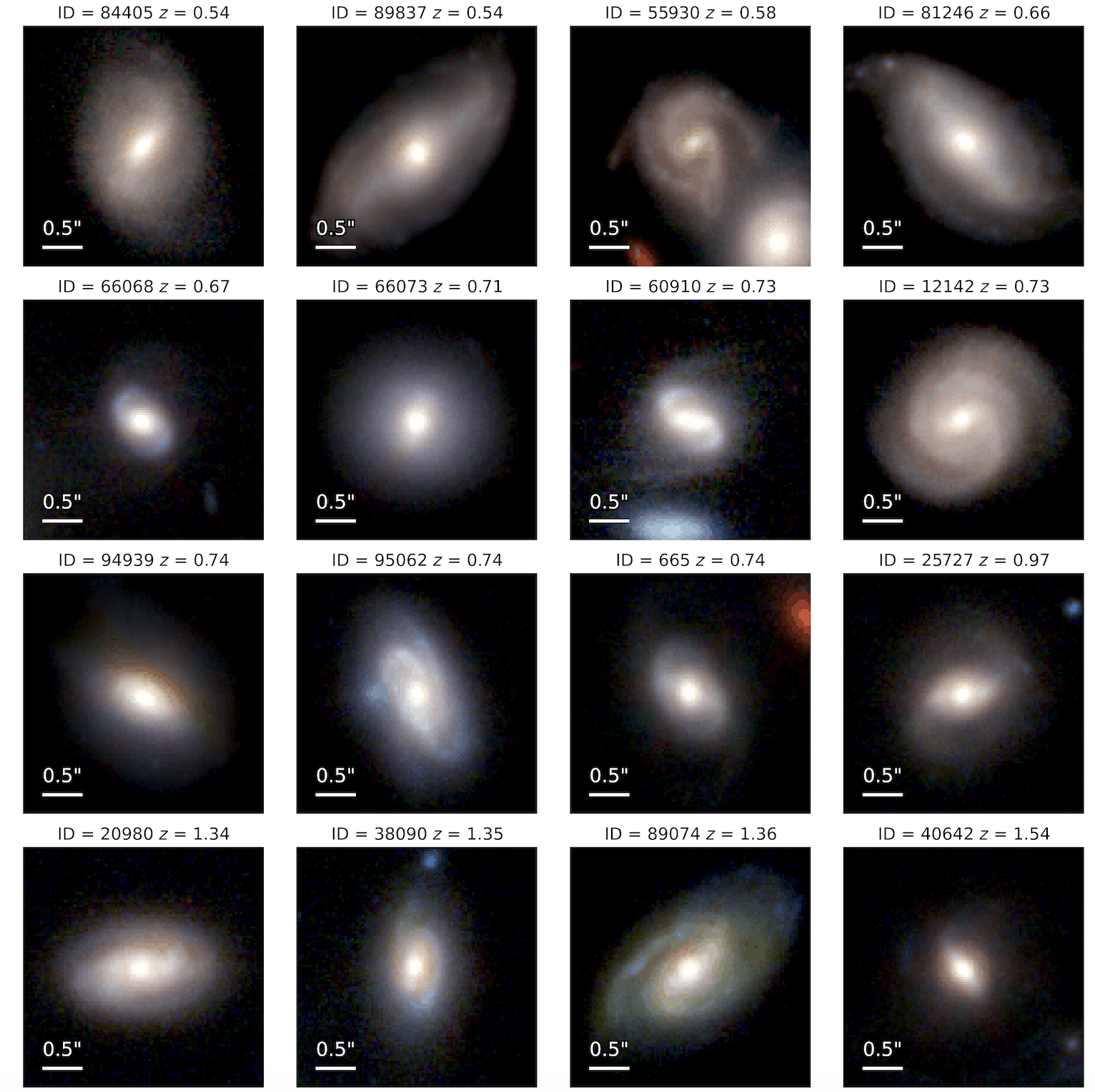}
         \caption{Selection of quiescent barred galaxies from the sSFR method (see Section~\ref{sec:quiescent id}). Quiescent disks exhibit smoother morphologies, redder colors, and little evidence of ongoing SF. The RGB images were constructed using \textit{JWST} NIRCam filters, with F115W mapped to the blue channel, F200W to green, and F444W to red.
}
\label{fig:quies_panel}
\end{figure*}

As described by \citet{Sherman2020}, several methods are commonly employed to identify quiescent galaxies including sSFR \citep[e.g., ][]{Fontanot2009}, distance from the main sequence \citep[e.g., ][]{Fang2018, Donnari2019}, and rest-frame colors \citep[e.g., ][]{Wuyts2007,Williams2009,Muzzin2013}. Each method introduces different advantages, assumptions, and limitations. 

Adopted widely across the literature, the sSFR selection method defines quiescent galaxies as those with sSFRs lower than a certain threshold, often $\mathrm{sSFR} < 10^{-11}\ \mathrm{yr^{-1}}$. This method has the advantage of being  easily applied and straightforward to compute using SED fitting outputs. However, the use of a constant sSFR threshold across a range of redshifts may not be physically motivated as it does not account for the evolution of the stellar mass-SFR relation with redshift \citep{Schreiber2015,Tomczak2016}. Some studies overcome this limitation by using a time-evolving sSFR limit \citep[e.g., ][]{Pacifici2016, Carnall2023}.

The main sequence selection method utilizes the sample of galaxies to define a main sequence in each redshift bin, and defines quiescent galaxies as those more than 1 dex below the main sequence \citep{Fang2018, Donnari2019}. This physically motivated method accounts for the evolution of the stellar mass-SFR relation with redshift \citep{Schreiber2015,Tomczak2016}; however, it requires a relatively robust sample to define an accurate main sequence and may vary depending on the SFR indicators used as well as the methods employed to derive stellar masses and SFRs. 

Rest-frame UVJ color selection uses the rest-frame U-V vs. V-J plane (or UVJ diagram) to define galaxies which fall in the top left area of the parameter space as quiescent \citep[e.g., ][]{Wuyts2007,Williams2009,Muzzin2013}. This method exploits the observed bimodality between quiescent and star-forming galaxies in rest-frame UVJ space and aims to break the degeneracy between dust reddening and age of the stellar population. However, classifications require precise rest-frame colors and can be impacted by dust, age, and metallicity. 

\citet{Sherman2020} found that consistent quiescent fractions were obtained using the aforementioned methods at 1.5 $<$ $z$ $<$ 3. For our main analysis, we identify quiescent galaxies using two methods. As is widely used in the literature, we first apply a constant sSFR threshold at $10^{-11}\ \mathrm{yr^{-1}}$ uniformly across all redshifts \citep{Fontanot2009}. Galaxies with $\mathrm{sSFR} < 10^{-11}\ \mathrm{yr^{-1}}$ are defined as quiescent (Figure~\ref{fig:sfr_mass}). We report the numbers of quiescent galaxies identified in each redshift bin: 69 (0.5 $\leq$ $z$ $<$ 1.0), 55 (1.0 $\leq$ $z$ $<$ 1.5), 44 (1.5 $\leq$ $z$ $\leq$ 2.0).

We also identify galaxies using the main sequence selection method. In this method, the main sequence
refers to the sequence of star-forming galaxies. Defining an accurate main sequence requires the stellar mass-SFR plane to be well populated across a wide range of stellar masses. Given the sparsity of data at high stellar masses in the stellar mass-SFR plane in the CEERS sample, we decided to use the published main sequence from \citet{Popesso2023}, which uses a wide set of studies to define the main sequence consistently across a range of redshifts. We utilize the main sequence parameterization presented in Equation 10 and the corresponding best-fit parameters reported in Table 2 of \citet{Popesso2023}. The midpoint of each redshift bin (0.5 $\leq$ $z$ $<$ 1.0, 1.0 $\leq$ $z$ $<$ 1.5, 1.5 $\leq$ $z$ $\leq$ 2.0) is adopted when applying the relations from \citet{Popesso2023}. The results are shown in Figure~\ref{fig:CEERS_MS}. Quiescent galaxies are defined as those with log SFR more than 1 dex below this main sequence relation (Figure~\ref{fig:CEERS_MS}). We report the numbers of quiescent galaxies identified in each redshift bin: 79 (0.5 $\leq$ $z$ $<$ 1.0), 72 (1.0 $\leq$ $z$ $<$ 1.5), 62 (1.5 $\leq$ $z$ $\leq$ 2.0). In Figure~\ref{fig:CEERS_MS}, the main sequence from \citet{Popesso2023} seems to lie above most CEERS data points; this issue may be tied to calibrations of SFRs. As a cross-check, we define the main sequence using CEERS data alone in the Appendix and verify that the key results of this paper based on the main sequence method remain unchanged, irrespective of whether we use the main sequence from CEERS or the one from \citet{Popesso2023}.

In the rest of this paper, we will compare the results on the quiescent bar fraction based on sSFR and main sequence selection methods. To illustrate the effectiveness of these classifications, we show a representative selection of barred galaxies identified as quiescent using the sSFR selection method (Figure~\ref{fig:quies_panel}). Visual inspection confirms that quiescent disks predominantly exhibit smooth morphologies, red colors, and little evidence of ongoing SF.

\begin{figure*}[!htb]
         \centering
         \includegraphics[width=1\textwidth]{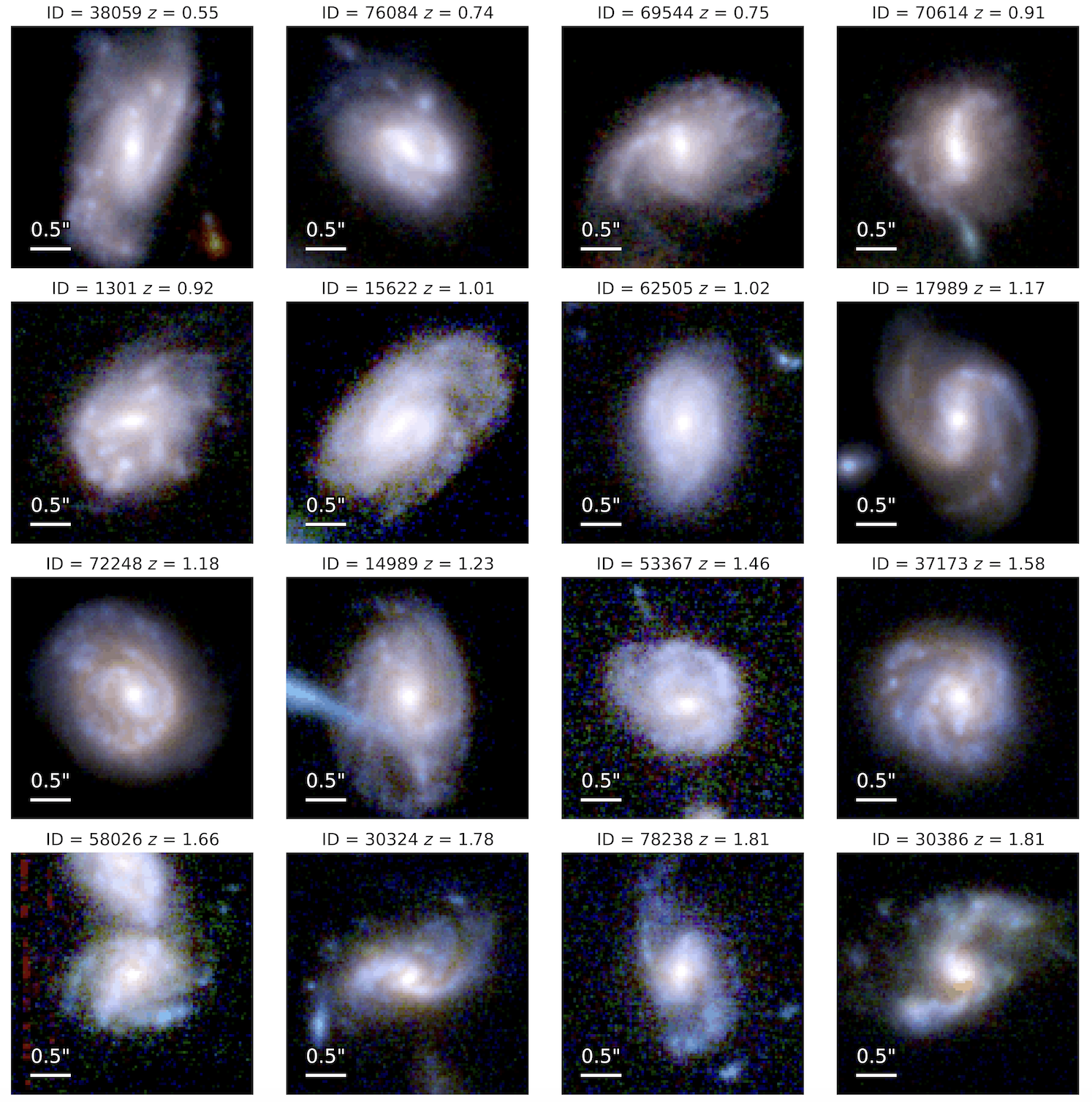}
         \caption{Selection of actively star-forming barred galaxies from the sSFR method (see Section~\ref{sec:starform id}). Actively star-forming disks exhibit clumpier morphologies, bluer colors, and evidence of ongoing SF. As in Figure~\ref{fig:quies_panel}, the RGB images were constructed using \textit{JWST} NIRCam filters, with F115W mapped to the blue channel, F200W to green, and F444W to red.
}
\label{fig:sf_panel}
\end{figure*} 

\subsection{Identifying Actively Star-Forming Galaxies}\label{sec:starform id}

We also explore in this paper the relationship between bars and actively star-forming galaxies. Unlike in the case of quiescent galaxies, there is no universally agreed-upon definition for this class of galaxies. We choose to identify this class of galaxies based on their sSFR, analogous to one of the methods we used to identify quiescent galaxies.

As illustrated in Figure~\ref{fig:sfr_mass}, we consider actively star-forming galaxies to have $\mathrm{sSFR} > 10^{-10}\ \mathrm{yr^{-1}}$, which is over an order of magnitude higher than the sSFR cut of $ < 10^{-11}\ \mathrm{yr^{-1}}$ used for quiescent galaxies. With this method, the numbers of star-forming galaxies identified in each redshift bin are 52 (0.5 $\leq$ $z$ $<$ 1.0), 143 (1.0 $\leq$ $z$ $<$ 1.5), and 108 (1.5 $\leq$ $z$ $\leq$ 2.0), respectively. 

As an illustration of our selection method, Figure~\ref{fig:sf_panel} shows a sample of barred galaxies identified as actively star-forming using the $\mathrm{sSFR} > 10^{-10}\ \mathrm{yr^{-1}}$ criterion. These barred disks exhibit bluer colors and clumpy features associated with active SF, unlike the quiescent barred galaxies which show smoother morphologies, redder colors, and little evidence of ongoing SF (Figure~\ref{fig:quies_panel}).

\subsection{Using COSMOS-Web Data as a Secondary Check on CEERS Results }\label{sec:COSMOS check}

The analyses in this paper are primarily performed using the CEERS survey and the bars identified therein by \citet{Guo2025}. However, in order to test the robustness of our results, we verify that key results hold if we use the sample of COSMOS-Web galaxies presented in \citet{HuertasCompany2025}. 
The large survey area of COSMOS-Web ($\sim$ 0.54 $\mathrm{deg}^2$ or $\sim$ 1920 $\mathrm{arcmin}^2$ vs. $\sim$ 90 $\mathrm{arcmin}^2$ in CEERS) also reduces the impact of cosmic variance and provides for more robust number statistics. COSMOS-Web provides NIRCam imaging in F115W, F150W, F277W, and F444W and MIRI observations in F770W, with NIRCam filters reaching a 5$\sigma$ depth ranging from 27.2 to 28.2, and the MIRI filter reaching a 5$\sigma$ depth of $\sim25.7$ \citep{Casey2023}.

We initially select massive galaxies with $M_\star > 10^{10}\,M_\odot$ to construct a parent sample comparable to that used in the CEERS analysis. This leads to a sample of 14,117 galaxies over $z \sim$ 0 to $z \sim$ 2.
 
In order to robustly measure the bar fraction, we first need to define a sample of moderately inclined disk galaxies from COSMOS-Web using the parameters provided in the catalog of \citet{HuertasCompany2025}. We identify disk-dominated galaxies using the morphological classification flag ``morph flag = 1'' defined in the neural network models of \citet{HuertasCompany2024}, where a supervised machine learning model is trained on visual morphology labels from the CANDELS survey \citep{Grogin2011,Koekemoer2011} and adapted to \textit{JWST} imaging via adversarial domain adaptation. For COSMOS-Web, the authors adopt the same architecture, input image size (32×32 pixels), and normalization, but the model is retrained from scratch using COSMOS-Web images as the target domain \citep{HuertasCompany2025}. To identify moderately inclined disk galaxies, we select disk-dominated systems with a low probability of being edge-on ($p_{edge-on}$ $<$ 0.5) based on probabilities derived from the Zoobot\footnote{\url{https://github.com/mwalmsley/zoobot}} deep learning model \citep{Walmsley2023}. Zoobot is a probabilistic deep learning foundation model for galaxy morphology classification trained on citizen science labels collected across several Galaxy Zoo campaigns \citep{Walmsley2023}.

From this sample of moderately inclined disk galaxies, we identify barred systems using the Zoobot bar probabilities. Following \citet{HuertasCompany2025}, we adopt $p_{bar}$ $>$ 0.5 as the threshold for a secure bar detection, where $p_{bar}$ is the mean probability a galaxy hosts a bar. Using stellar mass and SFR estimates derived with CIGALE \citep{Shuntov2025}, we identify quiescent and actively star-forming barred galaxies in COSMOS-Web using the sSFR and main sequence methods we applied for the CEERS sample (Sections~\ref{sec:quiescent id} and~\ref{sec:starform id}). 

The COSMOS-Web sample from \citet{HuertasCompany2025} has many advantages, including a large survey area, large sample size, robust number statistics, and reduced impact of cosmic variance. However, one limitation is that stellar bars are classified via neural networks trained on visual classifications and do not have the quantitative properties (e.g., maximum ellipticity, strength, semi-major axis) that we have in the CEERS sample from ellipse fitting. As mentioned earlier, these quantitative properties allow us to address systematic effects tied to bar and disk size, perform tailored apples-to-apples comparisons with simulations, and study the evolution of barred galaxies over cosmic time. Therefore, in this paper, we use both the COSMOS-Web and CEERS samples and leverage their complementary strengths.

\section{Results}\label{sec:results}

\subsection{Relationship between Barred Galaxies, SFR, sSFR, and Sérsic Index}\label{sec:host properties}

\begin{figure*}[!htb]
         \centering
         \includegraphics[width=1\textwidth]{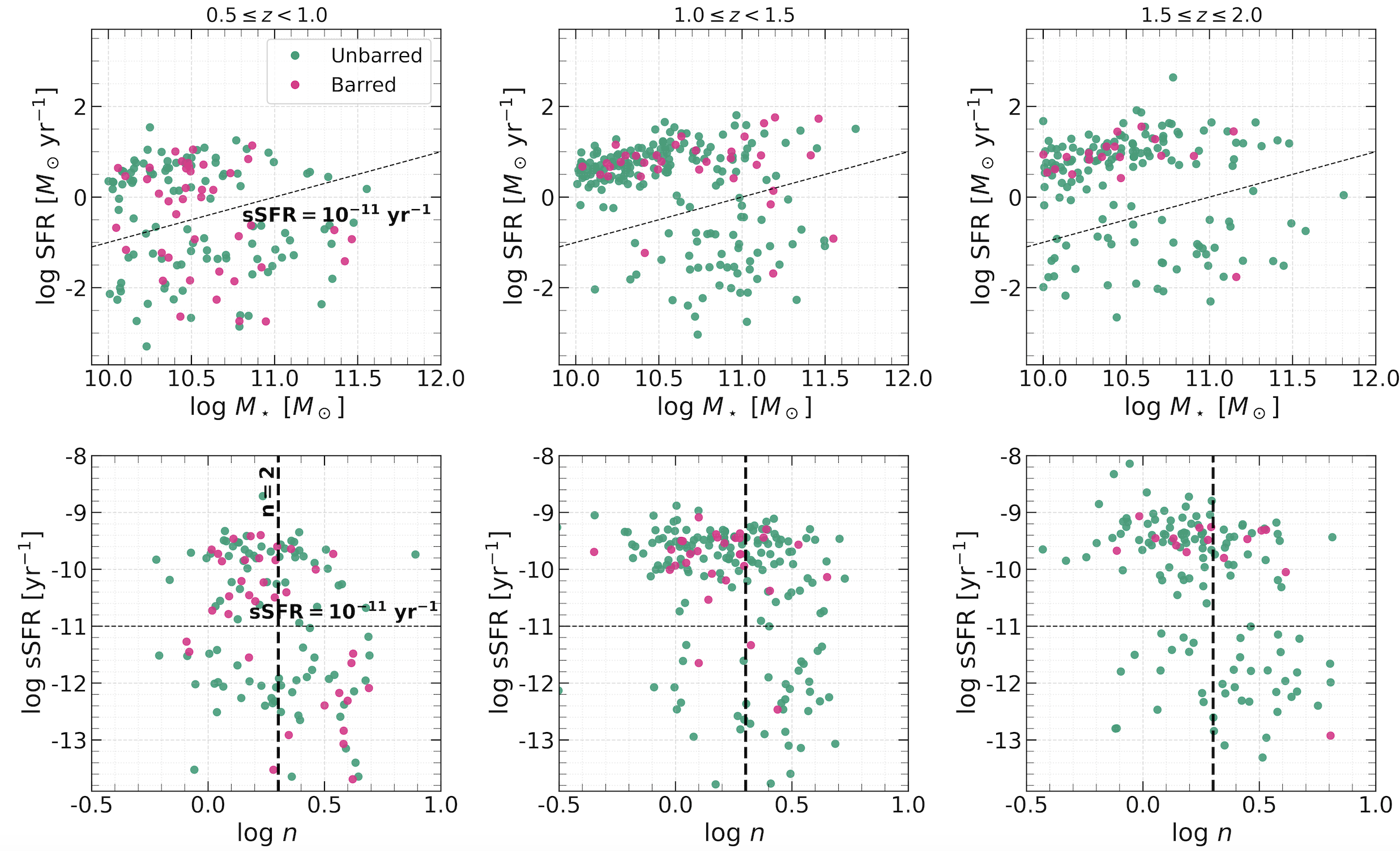}
         \caption{Distribution of disk galaxies in CEERS across three redshift bins: $0.5 \leq z < 1.0$, $1.0 \leq z < 1.5$, and $1.5 \leq z \leq 2.0$ (left to right). Barred and unbarred galaxies are shown as pink and green points, respectively.
         \textbf{Top}: $\log\,{\rm SFR}$ versus log stellar mass. The dashed line marks $\log\,{\rm sSFR} = -11~{\rm yr}^{-1}$. Barred galaxies evolve toward lower sSFR in the stellar mass-SFR plane.
         \textbf{Bottom}: $\log\,{\rm sSFR}$ versus log Sérsic index. The vertical dashed line at $n=2$ ($\log n \approx 0.3$) separates disk-dominated and bulge-dominated morphologies. At high redshift ($1.5\leq z<2$ and $1.0\leq z<1.5$) barred galaxy hosts predominantly have $n \leq 2$ ($\log(n) \leq0.3$), corresponding to disk-dominated galaxies that do not have well-developed bulges. However, in the lowest redshift bin ($0.5\leq z<1.0$), we see the emergence of barred galaxy hosts that have both higher $n$ ($ > 2$) and low sSFR.         
         }
    \label{fig:n_ssfr}
\end{figure*}

\begin{figure*}[!htb]
         \centering
         \includegraphics[width=1\textwidth]{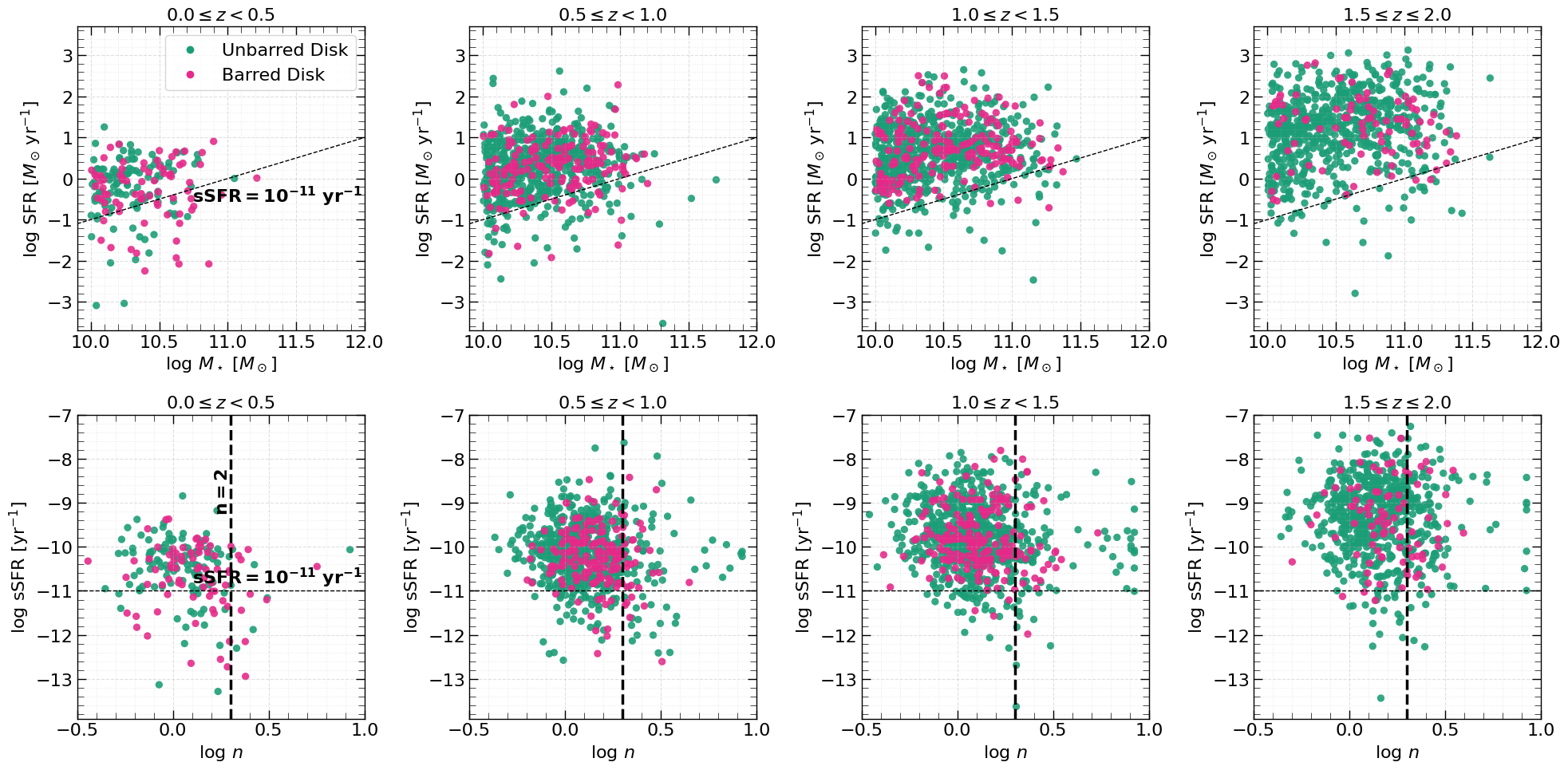}
         \caption{The same as Figure~\ref{fig:n_ssfr}, but for COSMOS-Web galaxies. As for CEERS galaxies, COSMOS-Web barred galaxies evolve toward lower sSFR in the stellar mass–SFR plane, and we see the emergence  of barred galaxy hosts that have both higher $n$ ($ > 2$) and low sSFR in the lowest redshift bin ($0\leq z<0.5$).      
         }
    \label{fig:n_ssfr_cos}
\end{figure*}

The relationship between barred galaxies, sSFR, and Sérsic index/bulge prominence has been studied at $z\sim0$ (e.g., \citealt{Weinzirl2009, Masters2010b, Masters2010a, Cheung2013, Gargiulo2022}). In this section we perform one of the very first explorations of these relationships out to higher redshifts, from $z\sim0$ to $z\sim2$.

Figures~\ref{fig:n_ssfr} and \ref{fig:n_ssfr_cos} show the distributions of disk galaxies in CEERS and COSMOS-Web in the $\log\,{\rm SFR}$ versus log stellar mass (top row), and $\log\,{\rm sSFR}$ versus log Sérsic index (bottom row) planes, split into multiple redshift bins ($0 \leq z < 0.5$ (COSMOS-Web only), $0.5 \leq z < 1.0$, $1.0 \leq z < 1.5$, and $1.5 \leq z \leq 2.0$). Barred and unbarred galaxies are shown as magenta and green points, respectively. As mentioned previously, CEERS Sérsic indices are taken from \citet{McGrath2026}, who derived structural parameters using \texttt{GALFIT} single-component Sérsic fits. COSMOS-Web Sérsic indices in \citet{HuertasCompany2025} are taken from \citet{Shuntov2025}, who derived structural parameters using \texttt{SE++} profile fitting. The dashed diagonal lines at $\log\,{\rm sSFR} = -11~{\rm yr}^{-1}$ mark the threshold below which we consider galaxies to be quiescent systems, while the vertical dashed lines at $n=2$ ($\log n \approx 0.3$) separate disk-dominated and bulge-dominated morphologies.

We first explore the relationship between bars, SFR, and sSFR. \citet{Cheung2013} found that, at $z\sim0$, the likelihood of a galaxy hosting a bar is anticorrelated with sSFR, regardless of stellar mass or bulge prominence. The top rows of Figure~\ref{fig:n_ssfr} and Figure~\ref{fig:n_ssfr_cos} show that at high redshift ($z\sim1$--$2$) barred galaxies tend to have predominantly high sSFRs while in the lowest redshift bins ($0.5\leq z <1$ and $0\leq z<0.5$), a large number of barred galaxies occupy the low sSFR regime. This evolution toward lower sSFR is evident examining the host galaxies of bars as a function of stellar mass (top row) and Sérsic index (bottom row). 

Next, we explore the relationship between bars and the Sérsic index of their hosts. Among $z\sim0$ galaxies, \citet{Masters2010a} report that there is a clear increase in the bar fraction with redder (g-r) colors, decreased luminosity, and in galaxies with more prominent bulges, to the extent that over half of the red, bulge-dominated disk galaxies in their sample possess a bar. In our study, the bottom rows of Figure~\ref{fig:n_ssfr} and Figure~\ref{fig:n_ssfr_cos} show that at high redshift ($1.5\leq z \le2$ and $1.0\leq z<1.5$) barred galaxy hosts predominantly have $n \leq 2$ ($\log(n) \leq0.3$), corresponding to disk-dominated galaxies that do not have well-developed bulges. However, in the lowest redshift bins ($0.5\leq z <1$ and $0\leq z<0.5$), we see the emergence of barred galaxy hosts that have both higher $n$ ($ > 2$) and low sSFR in the lower right quadrant.  Specifically, the percentage of barred galaxies in CEERS in this lower right quadrant rises from 6.7\% in the highest redshift bin to 27\% in the lowest redshift bin. Similar results are found in COSMOS-Web, with the percentage of barred galaxies in the lower right quadrant rising from 0\% in the highest redshift bin to 6.5\% in the lowest redshift bin. This trend, combined with the findings of \citet{Masters2010a}, may be consistent with bar-driven secular evolutionary scenarios where repeated bar-driven gas inflows lead to high CN SFRs, which build compact CN disks called pseudobulges, while transitioning galaxies to a quiescent phase. We discuss these different phases of evolution in a barred galaxy in more detail in Section~\ref{sec:scen1} and the schematic figure therein. One limitation of this analysis is that we rely on single-Sérsic profile fits rather than full multi-component decompositions. As a result, it is difficult to disentangle the relative contributions of disks, bulges, bars, and other central structures to the observed light distribution. While higher Sérsic indices ($n \gtrsim 2$) generally indicate a more centrally concentrated light profile, such concentrations do not necessarily correspond to a bulge. Instead, they may arise from a variety of structures, including pseudobulges, classical bulges, compact central starbursts, certain types of bars, and other components that contribute significantly to the central light distribution \citep[e.g.,][]{Gadotti2008,Weinzirl2009}. Consequently, the Sérsic indices presented here should be interpreted primarily as measures of central light concentration rather than direct measurements of bulge prominence.

\subsection{Contribution of Quiescent and Actively Star-Forming Galaxies to the Bar Fraction}\label{sec:SF Q contribution}

\begin{figure*}[!htb]
    \centering
    \begin{subfigure}{0.49\textwidth}
        \centering
        \includegraphics[width=\linewidth]{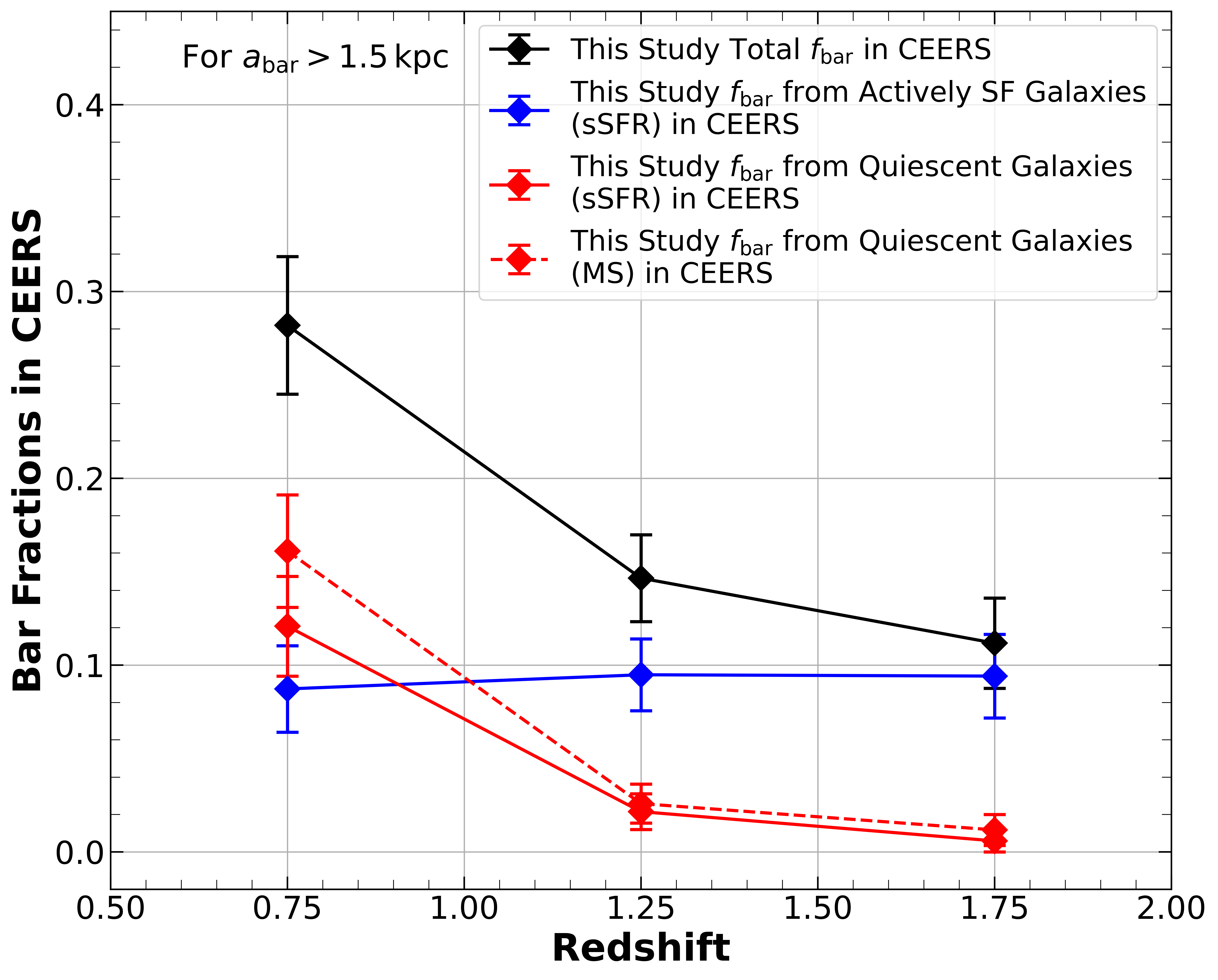}
    \end{subfigure}
    \hfill
    \begin{subfigure}{0.49\textwidth}
        \centering
        \includegraphics[width=\linewidth]{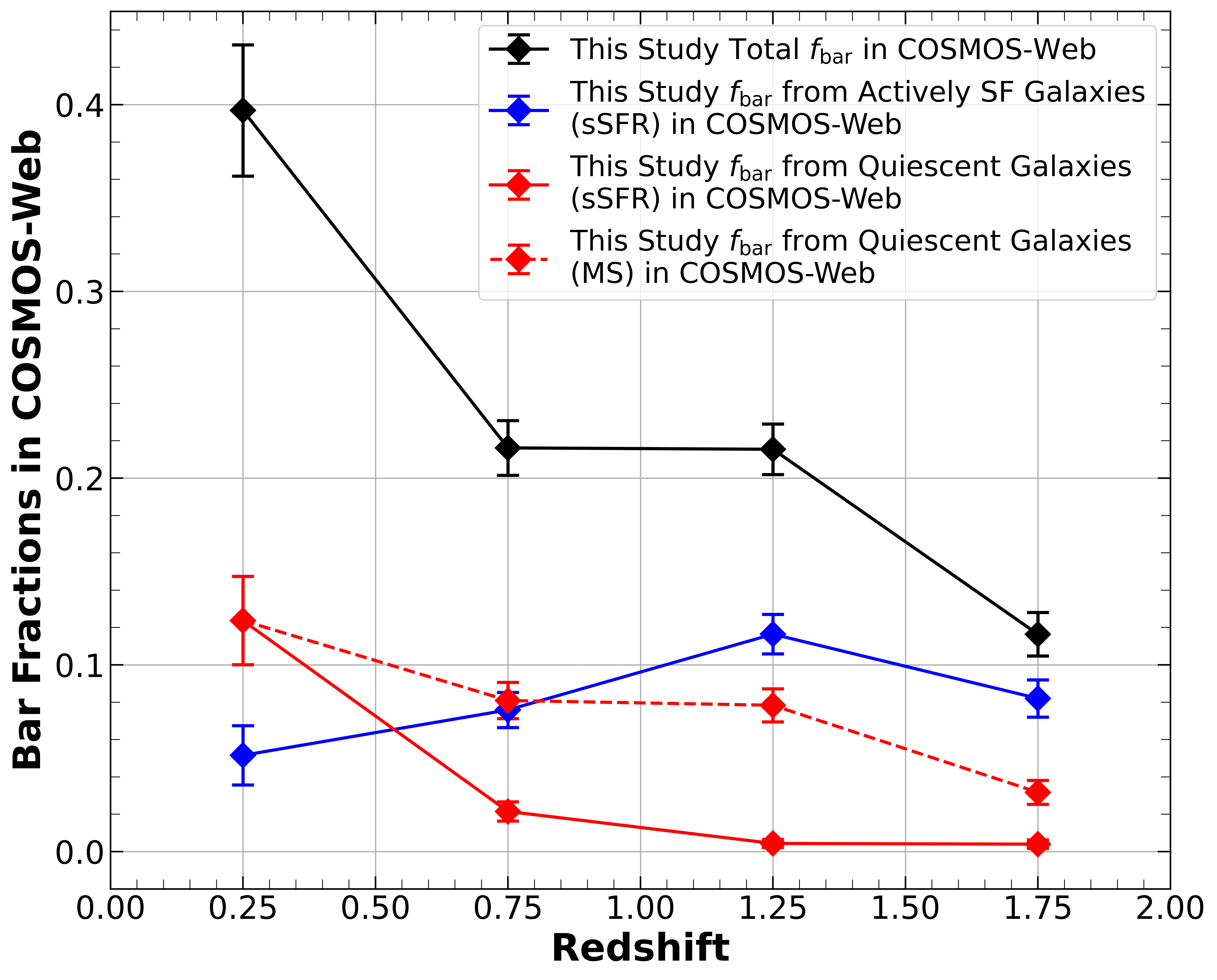}

    \end{subfigure}
    \caption{\textbf{Left}: The total bar fraction from \citet{Guo2025} based on ellipse fits is shown in black  for bars with semi-major axis $a_{\mathrm{bar}}$ $>$ 1.5 kpc that can be robustly detected via ellipse fits. Contributions from quiescent barred galaxies are shown in red (solid and dashed lines for sSFR and main sequence selection methods respectively), and from actively star-forming galaxies in blue. Barred quiescent galaxies contribute to less than $\sim$8\% (1\%/12\%) of the bar fraction at $z \sim$ 1.75 and as much as $\sim$43\% (12\%/28\%) by $z \sim$ 0.75, while barred actively star-forming galaxies contribute $\sim$75\% (9\%/12\%) of the bar fraction at $z \sim$ 1.75 and only $\sim$28\% (8\%/28\%) by $z \sim$ 0.75. \textbf{Right}: Same as the left panel, but for COSMOS-Web data. Barred quiescent galaxies contribute $\sim$4\% (0.5\%/12\%) to the bar fraction at $z \sim$ 1.75 and as much as $\sim$33\% (13\%/40\%) by $z \sim$ 0.25 while barred actively star-forming galaxies contribute $\sim$67\% (8\%/12\%) of the bar fraction at $z \sim$ 1.75 and only $\sim$12\% (5\%/40\%) by $z \sim$ 0.25.}
    \label{fig:contribute_CEERS}
\end{figure*} 

Using CEERS data, \citet{Guo2025} derived the bar fraction from $z \sim$ 0.5 to $z \sim$ 4 using visual classifications and ellipse fits, with both methods yielding similar results. They report the bar fraction rises from $6.4^{+3.4}_{-2.3}\%$ at $z \sim$ 3.5 to $28.4^{+3.8}_{-3.6}\%$ at $z \sim$ 0.75, and note that the results apply to bars with projected semi-major axis lengths $a_{\mathrm{bar}}$ $>$ 1.5 kpc, which can be resolved and robustly detected by ellipse fits. 
 
The left panel of Figure~\ref{fig:contribute_CEERS} shows the CEERS bar fraction $f_{bar}$ from \citet{Guo2025} based on ellipse fits. 
The right panel of Figure~\ref{fig:contribute_CEERS} shows the bar fraction $f_{bar}$ from COSMOS-Web data \citep{HuertasCompany2025} 
based on neural network classifications (see Section~\ref{sec:COSMOS check}). We also show the contributions to $f_{bar}$ from quiescent galaxies (i.e., the fraction of moderately inclined disks that are both quiescent and barred) identified using the sSFR method and main sequence methods (Section~\ref{sec:quiescent id}), and from actively star-forming galaxies identified via their sSFR (Section~\ref{sec:starform id}). The error bars represent the 1$\sigma$ statistical binomial uncertainties on the measured bar fractions. All subsequent errors are computed in the same manner unless otherwise noted.

CEERS and COSMOS-Web data show that the fractional contribution of barred quiescent galaxies to the bar fraction rises significantly from $z \sim$ 2 to $z \sim$ 0 (Figure~\ref{fig:contribute_CEERS}). With the sSFR method of identifying quiescent galaxies, in CEERS barred quiescent galaxies contribute to less than $\sim$8\% (1\%/12\%) of the bar fraction at $z \sim$ 1.75 and as much as $\sim$43\% (12\%/28\%) by $z \sim$ 0.75. In COSMOS-Web, barred quiescent galaxies contribute $\sim$4\% (0.5\%/12\%) to the bar fraction at $z \sim$ 1.75 and as much as $\sim$33\% (13\%/40\%) by $z \sim$ 0.25.
 
In contrast, the fractional contribution of barred actively star-forming galaxies to the bar fraction falls over time from $z \sim$ 2 to $z \sim$ 0 (Figure~\ref{fig:contribute_CEERS}). In CEERS, barred actively star-forming galaxies contribute $\sim$75\% (9\%/12\%) of the bar fraction at $z \sim$ 1.75 and only $\sim$28\% (8\%/28\%) by $z \sim$ 0.75. In COSMOS-Web, barred actively star-forming galaxies contribute $\sim$67\% (8\%/12\%) of the bar fraction at $z \sim$ 1.75 and only $\sim$12\% (5\%/40\%) by $z \sim$ 0.25.
 
We will discuss possible explanations for these trends in Section~\ref{sec:discussion}.

\subsection{The Bar Fraction in Quiescent and Actively Star-Forming Galaxies}\label{sec:quiescent SF fbar}

Another way to gain insight into the impact of bars on SF activity is to explore the evolution of the bar fraction among quiescent and actively star-forming galaxies. At $z \sim$ 0, \citet{Masters2010b} report a large optical bar fraction in red spirals (70\% $\pm$ 5\% versus 27\% $\pm$ 5\% in blue spirals) and suggest that the cessation of SF and bar instabilities in spirals are strongly correlated. However, no such explorations have been conducted at higher redshift; in Sections~\ref{sec:quiescent SF fbar CEERS COSMOS} to \ref{sec:theory}, we perform the first systematic exploration of how the bar fraction varies over time among quiescent and actively star-forming galaxies out to $z \sim 2$.

\subsubsection{The Bar Fraction in Quiescent and Actively Star-Forming Galaxies from CEERS and COSMOS-Web}\label{sec:quiescent SF fbar CEERS COSMOS}

\begin{deluxetable*}{c|cc|cc}

\tablecaption{Bar Fraction Among Quiescent Galaxies ($F_{\mathrm{Q\_bar}}$)\label{tab:bar_fraction_quiescent}}

\tablehead{
\colhead{}  & \multicolumn{2}{c}{CEERS} & \multicolumn{2}{c}{COSMOS-Web} \\
\colhead{Redshift Bin} & \colhead{$F_{\mathrm{Q\_bar}}$ (sSFR)} & \colhead{$F_{\mathrm{Q\_bar}}$ (MS $-1$ dex)} & \colhead{$F_{\mathrm{Q\_bar}}$ (sSFR)} & \colhead{$F_{\mathrm{Q\_bar}}$ (MS $-1$ dex)}
}

\startdata
0.0--0.5 & - & -  & 54.5\% $\pm$ 7.5\% (24/44) & 46.2\% $\pm$ 6.9\% (24/52) \\
0.5--1.0 & 26.1\% $\pm$ 5.3\% (18/69) & 30.4\% $\pm$ 5.2\% (24/79) &  22.4\% $\pm$ 4.8\% (17/76) & 23.0\% $\pm$ 2.5\% (64/278) \\
1.0--1.5 & 9.1\% $\pm$ 3.9\% (5/55) & 8.3\% $\pm$ 3.3\% (6/72) & 8.7\% $\pm$ 4.2\% (4/46) & 21.1\% $\pm$ 2.2\% (72/342) \\
1.5--2.0 & 2.3\% $\pm$ 2.3\% (1/44) & 3.2\% $\pm$ 2.2\% (2/62) &  9.7\% $\pm$ 5.3\% (3/31) & 13.5\% $\pm$ 2.6\% (24/178) \\
\enddata

\tablenotetext{}{Bar fractions are shown as percentages, with the number of barred galaxies over the total number of moderately inclined quiescent disks in parentheses. Quiescent galaxies are defined either as having a $\mathrm{sSFR} < 10^{-11}\,\mathrm{yr^{-1}}$ or as having SFRs below the main sequence -1 dex curve.}

\end{deluxetable*}
\begin{figure*}[!htb]
    \centering
    \begin{subfigure}{0.49\textwidth}
        \centering
        \includegraphics[width=\linewidth]{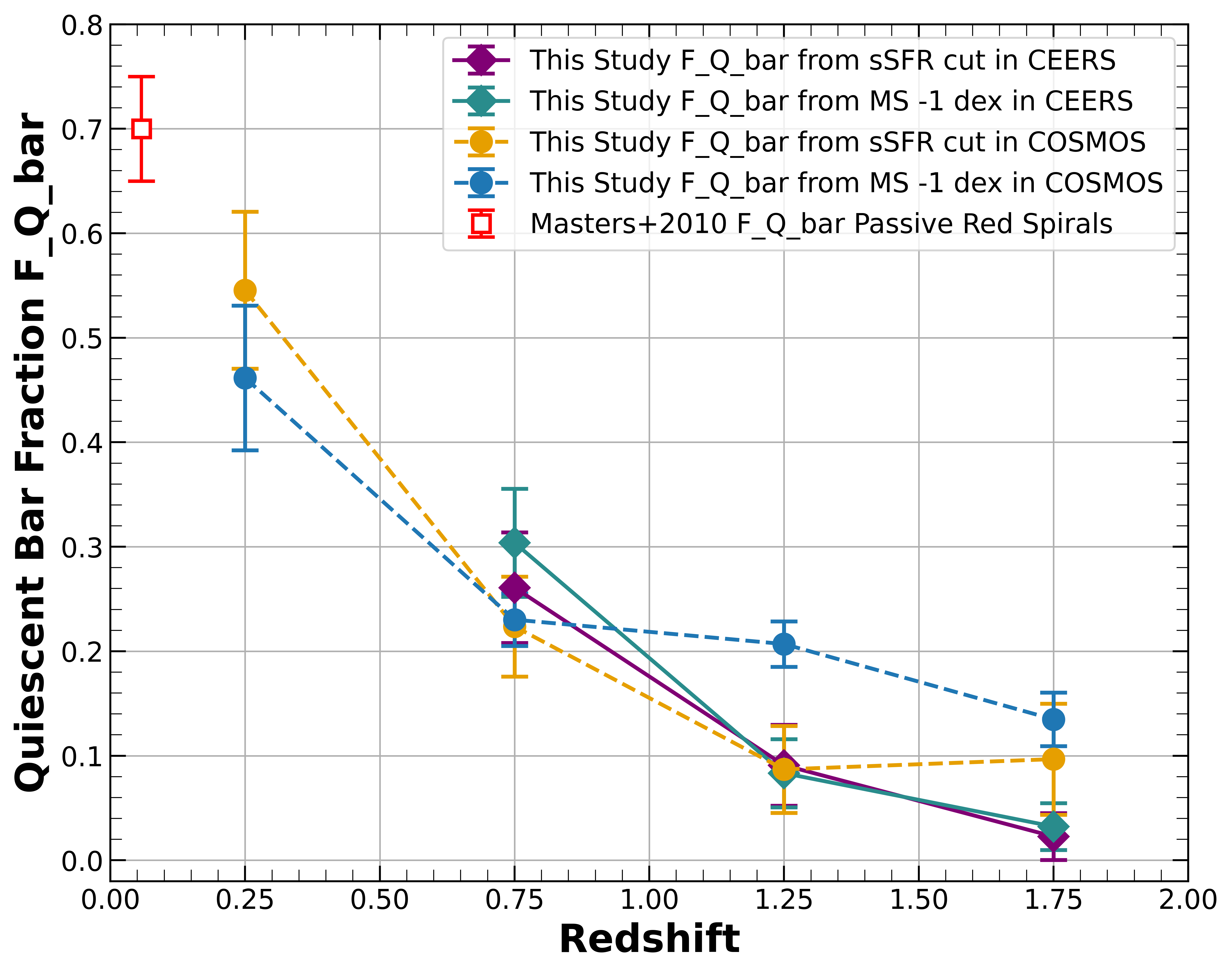}

    \end{subfigure}
    \hfill
    \begin{subfigure}{0.49\textwidth}
        \centering
        \includegraphics[width=\linewidth]{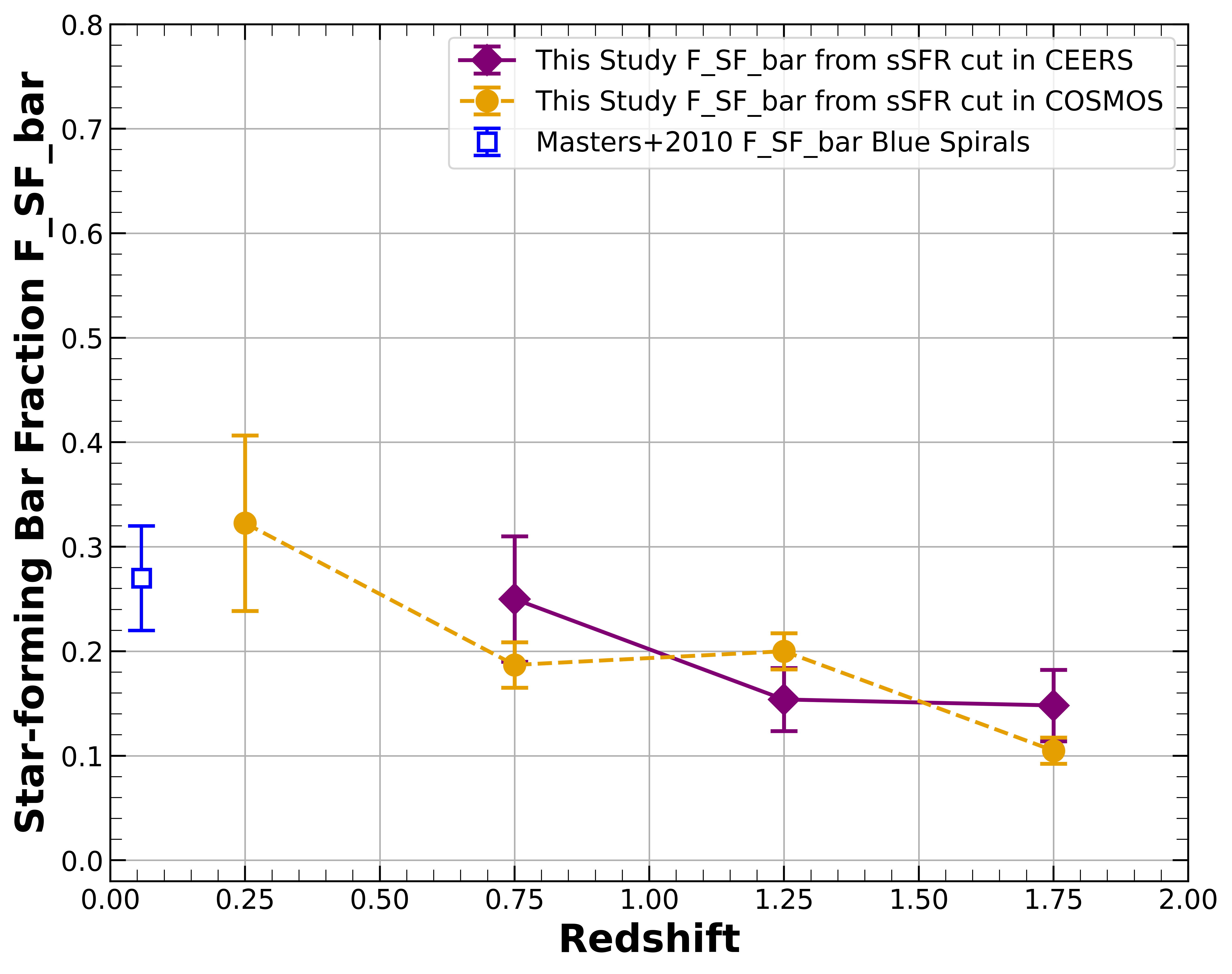}
    \end{subfigure}
    \caption{\textbf{Left}: The observed bar fraction among quiescent disk galaxies with $M_\star > 10^{10}\,M_\odot$ ($F_{\mathrm{Q\_bar}}$) is plotted from $z \sim 2$ to $z \sim 0$ based on CEERS and COSMOS-Web data. The value of $F_{\mathrm{Q\_bar}}$ is plotted against the midpoint of each redshift bin. Quiescent galaxies are identified using both the sSFR and main sequence selection methods (Section~\ref{sec:quiescent id}). The quiescent bar fraction rises steeply over the last 10 Gyr ($z \sim 2$ to $z \sim 0$) from $\sim 2.3\%\pm22.3\%$ at $z \sim 1.75$ to $\sim 54.5\%\pm7.5\%$ at $z \sim 0.25$. For reference, we also plot the bar fraction ($70\%\pm5\%$) among $z \sim 0$ red spirals (quiescent disks) in SDSS reported by \citet{Masters2010b}. \textbf{Right}: Same as the left panel, but plotting the fraction of actively star-forming disk galaxies that are barred ($F_{\mathrm{SF\_bar}}$). Actively star-forming galaxies are identified using a sSFR cut (Section~\ref{sec:starform id}). $F_{\mathrm{SF\_bar}}$ rises from $\sim10.5\%\pm 1.3\%$ at $z \sim 1.75$ to $\sim 32.3\%\pm 8.4\%$ by $z \sim 0.25$. While $F_{\mathrm{SF\_bar}}$ is higher than the quiescent bar fraction ($F_{\mathrm{Q\_bar}}$) at high redshift ($z \sim 2$), it increases at a much shallower rate with time and by $z \sim 0.25$ it is superseded by a higher $F_{\mathrm{Q\_bar}}$. For reference, we also plot the bar fraction ($27\%\pm5\%$) among $z \sim 0$ blue spirals in SDSS reported by \citet{Masters2010b}.}
\label{fig:F_Q_bar}
\end{figure*}

In this section, we explore the quiescent bar fraction ($F_{\mathrm{Q\_bar}}$) based on CEERS and COSMOS-Web data. The following steps are adopted to calculate the quiescent bar fraction. First, quiescent galaxies are identified using both the sSFR selection method and the main sequence selection method (Section~\ref{sec:quiescent id}). Among these quiescent galaxies, the bar fraction $F_{\mathrm{Q\_bar}}$ is calculated from a sample of moderately inclined disk galaxies using the equation below:
\begin{equation}
    f_{\mathrm{bar}} = \frac{N_{\mathrm{bar}}}{N_{\mathrm{disk}}}
    \label{eq:barfrac}
\end{equation}
where $N_{\mathrm{bar}}$ is the number of barred galaxies identified in a sample of moderately inclined disks (in this case quiescent moderately inclined disks) with total sample size $N_{\mathrm{disk}}$.

For CEERS, Sections~\ref{sec:vis class CEERS} to \ref{sec:bar id CEERS} outline how moderately inclined disks and bars are identified from the study by \citet{Guo2025}. \citet{Guo2025} identified bars in CEERS using visual classification and ellipse fitting methods. In this work, we adopt the bar classifications from ellipse fits and note that the results apply to bars with projected semi-major axis lengths $a_{\mathrm{bar}} > 1.5\,\mathrm{kpc}$, which can be resolved and robustly detected by ellipse fitting. For the COSMOS-Web data, we use the neural network classifications of disks and bars from \citet{HuertasCompany2025}, as described in Section~\ref{sec:COSMOS check}.

The results are shown in Table~\ref{tab:bar_fraction_quiescent} and the left panel of Figure~\ref{fig:F_Q_bar}. This panel shows the quiescent bar fraction ($F_{\mathrm{Q\_bar}}$) from $z \sim 2$ to $z \sim 0$ based on CEERS and COSMOS-Web data. The quiescent bar fraction is plotted against the midpoint of each redshift bin. The CEERS data show a steep increase in the quiescent bar fraction from approximately $2.3\%\pm2.3\%$ at $z \sim 1.75$ to $26.1\%\pm 5.3\%$ by $z \sim 0.75$. The COSMOS-Web data are consistent with this steep rise and indicate that this trend continues to $z \sim 0.25$. Furthermore, extrapolating this trend to $z \sim 0$ yields a quiescent bar fraction of $\sim 70\%\pm5\%$, consistent with the bar fraction of red spirals (quiescent disks) in SDSS reported by \citet{Masters2010b}, based on visual classifications from the Galaxy Zoo project. Taken together, these empirical analyses from different surveys show that \textit{the observed fraction of bars in quiescent disk galaxies rises steeply over the last $\sim10$ Gyr ($z \sim 2$ to $z \sim 0$), from $\sim 2.3\%\pm2.3\%$ at $z \sim 1.75$ to $\sim 54.5\%\pm7.5\%$ at $z \sim 0.25$, reaching $\sim 70\%\pm5\%$ at $z \sim 0$}. To our knowledge, this is the first time this trend has been identified. These findings align with our results in Section~\ref{sec:host properties} where we discuss the barred and unbarred populations in quiescent and actively star-forming galaxies and find that the number of quiescent barred galaxies increases significantly from $z \sim2$ to $z \sim0.5$. The significance of this result will be discussed in Section~\ref{sec:discussion} after we further assess its robustness in Section~\ref{sec:disk sizes CEERS}.

For comparison, the right panel of Figure~\ref{fig:F_Q_bar} shows the fraction of actively star-forming galaxies that are barred ($F_{\mathrm{SF\_bar}}$) from $z \sim 2$ to $z \sim 0.5$ based on CEERS and COSMOS-Web data. Actively star-forming galaxies are identified using the sSFR cut described in Section~\ref{sec:starform id}. Among these galaxies, the bar fraction $F_{\mathrm{SF\_bar}}$ is calculated from a sample of moderately inclined disk galaxies using equation~\ref{eq:barfrac}. We find that the actively star-forming bar fraction, $F_{\mathrm{SF\_bar}}$, rises from $\sim10.5\%\pm 1.3\%$ at $z \sim 1.75$ to $\sim 32.3\%\pm 8.4\%$ by $z \sim 0.25$. \textit{While $F_{\mathrm{SF\_bar}}$ is higher than the quiescent bar fraction ($F_{\mathrm{Q\_bar}}$) at high redshift ($z \sim 1.75$), it increases at a much shallower rate with time and is surpassed by $F_{\mathrm{Q\_bar}}$ by $z \sim 0.25$}. For reference, we also plot the bar fraction ($27\%\pm5\%$) among $z \sim 0$ blue spirals in SDSS reported by \citet{Masters2010b}. We discuss possible explanations for these trends in Section~\ref{sec:discussion}.

\subsubsection{Exploring Systematic Effects of Disk Sizes on the Quiescent Bar Fraction in CEERS}\label{sec:disk sizes CEERS}

\begin{figure}
    \centering
    \includegraphics[width=0.9\linewidth]{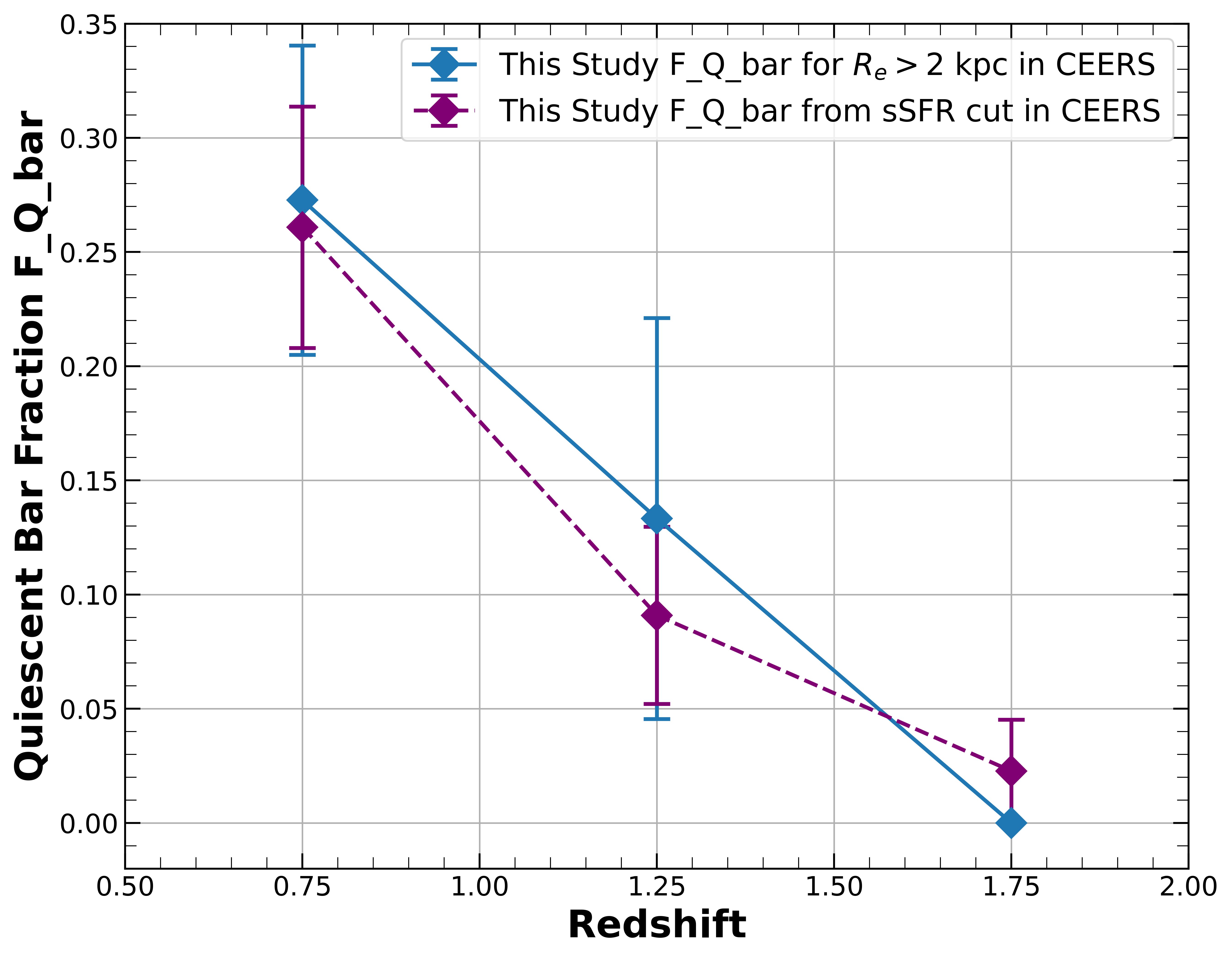}
    \caption{Overall quiescent bar fraction (dashed purple line) and quiescent bar fraction computed after excluding ultra-compact disks ($R_e < 2$ kpc) (solid blue line) in CEERS. The steep rise in the quiescent bar fraction observed remains unchanged, confirming that the results are not driven by a dominant population of ultra-compact quiescent disk galaxies at higher redshifts.}
    \label{fig:large_fbar}
\end{figure}

Many massive quiescent galaxies at $z >$ 1.5 tend to be more compact relative to star-forming galaxies of similar stellar masses at $z >$ 1.5, and relative to quiescent galaxies at $z\sim$ 0 \citep[e.g., ][]{Daddi2005, Weinzirl2011, VanderWel2014, Barro2013}. These systems, nicknamed ``red nuggets'', include massive ($M_\star \gtrsim 10^{10.5}\,M_\odot$) systems with effective radii of only $\sim$ 1 kpc, nearly an order of magnitude smaller than present-day early-type galaxies. Given the increased compactness of quiescent galaxies at $z >$ 1.5, we must ask whether the observed trend of a rising quiescent bar fraction $F_{\mathrm{Q\_bar}}$ from $z \sim$ 2 to $z \sim$ 0.5 (left panel of Figure~\ref{fig:F_Q_bar}) could be due to our inability to resolve and robustly detect bars in the large population of compact quiescent disks at $z >$ 1.5. Specifically, in extremely compact quiescent disks at $z >$ 1.5, the semi-major axis of any potential bar would be near or below the limit of 1.5 kpc ($2$ $\times$ the PSF FWHM of $\sim$ 0.08$"$ in F200W) required to robustly detect bars via ellipse fits. As such, although bars may exist in these systems, we may be unable to resolve and/or robustly detect them. In this section, we quantitatively explore the possible impacts of this systematic effect on our observed evolution of the quiescent bar fraction.

To determine whether the observed evolution in bar fraction in quiescent disk galaxies is driven by the compactness of quiescent disks at $z > 1.5$, we test the impact of disk sizes by excluding ultra-compact disks with $R_e < 2$ kpc using $R_e$ measurements in F356W derived by \citet{McGrath2026}, and recompute the bar fraction in quiescent disk galaxies. As we have shown that the sSFR and main sequence selection methods for identifying quiescent galaxies produce the same results in the quiescent bar fraction, we use the sSFR-selected quiescent galaxies in this section as our sample. Figure~\ref{fig:large_fbar} shows the quiescent bar fraction ($F_{\mathrm{Q\_bar}}$) computed with and without ultra-compact disks. It shows that even when ultra-compact disks are removed, we still see the overall trend of the quiescent bar fraction ($F_{\mathrm{Q\_bar}}$) rising sharply from $z \sim$ 2 to $z \sim$ 0.5. This demonstrates that our results on the quiescent bar fraction are fairly robust to biases stemming from the inability to resolve/detect bars in compact quiescent disks.

\begin{figure*}[!htb]
    \centering
    \begin{subfigure}{0.49\textwidth}
        \centering
        \includegraphics[width=\linewidth]{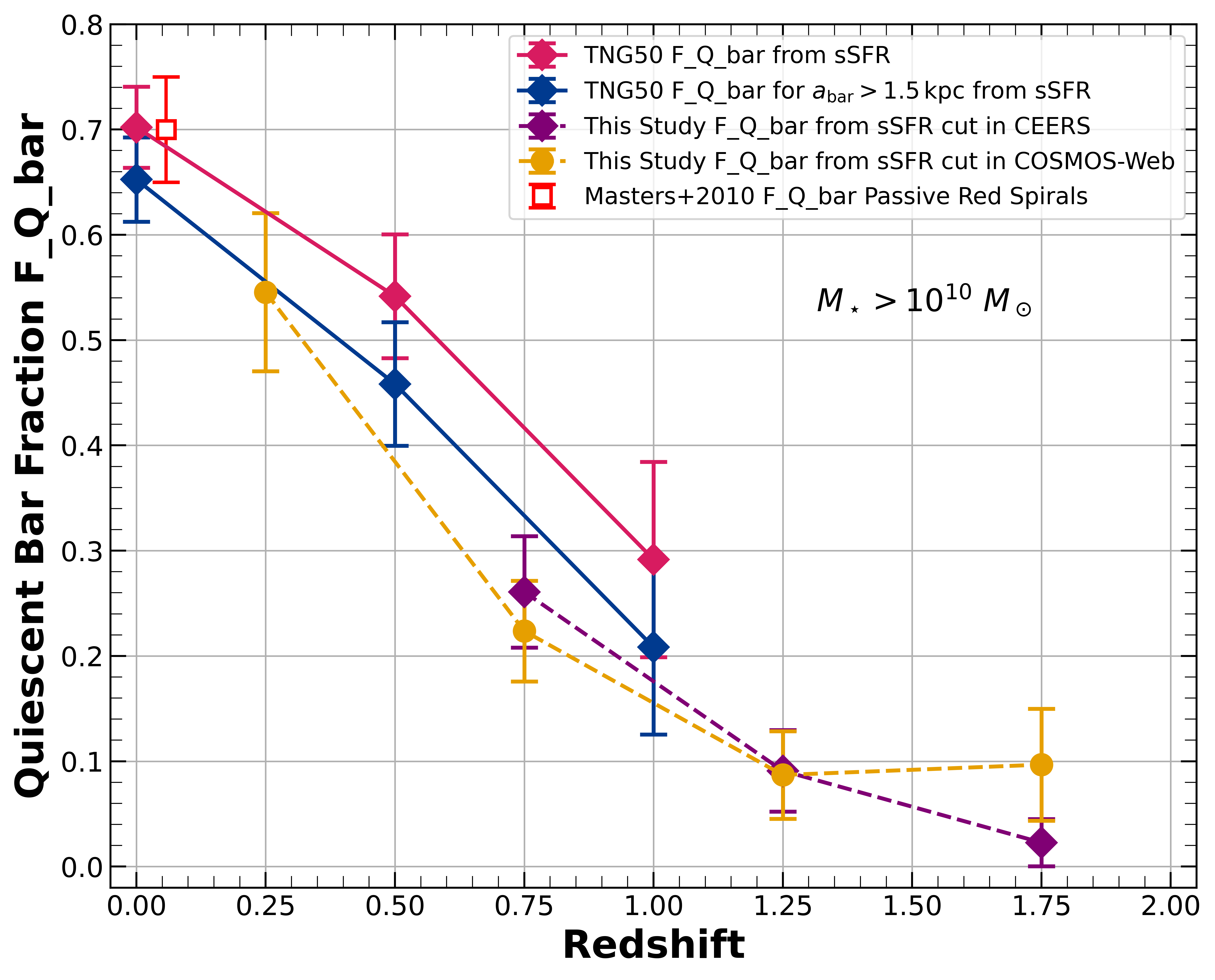}
       \end{subfigure}
    \hfill
    \begin{subfigure}{0.49\textwidth}
        \centering
        \includegraphics[width=\linewidth]{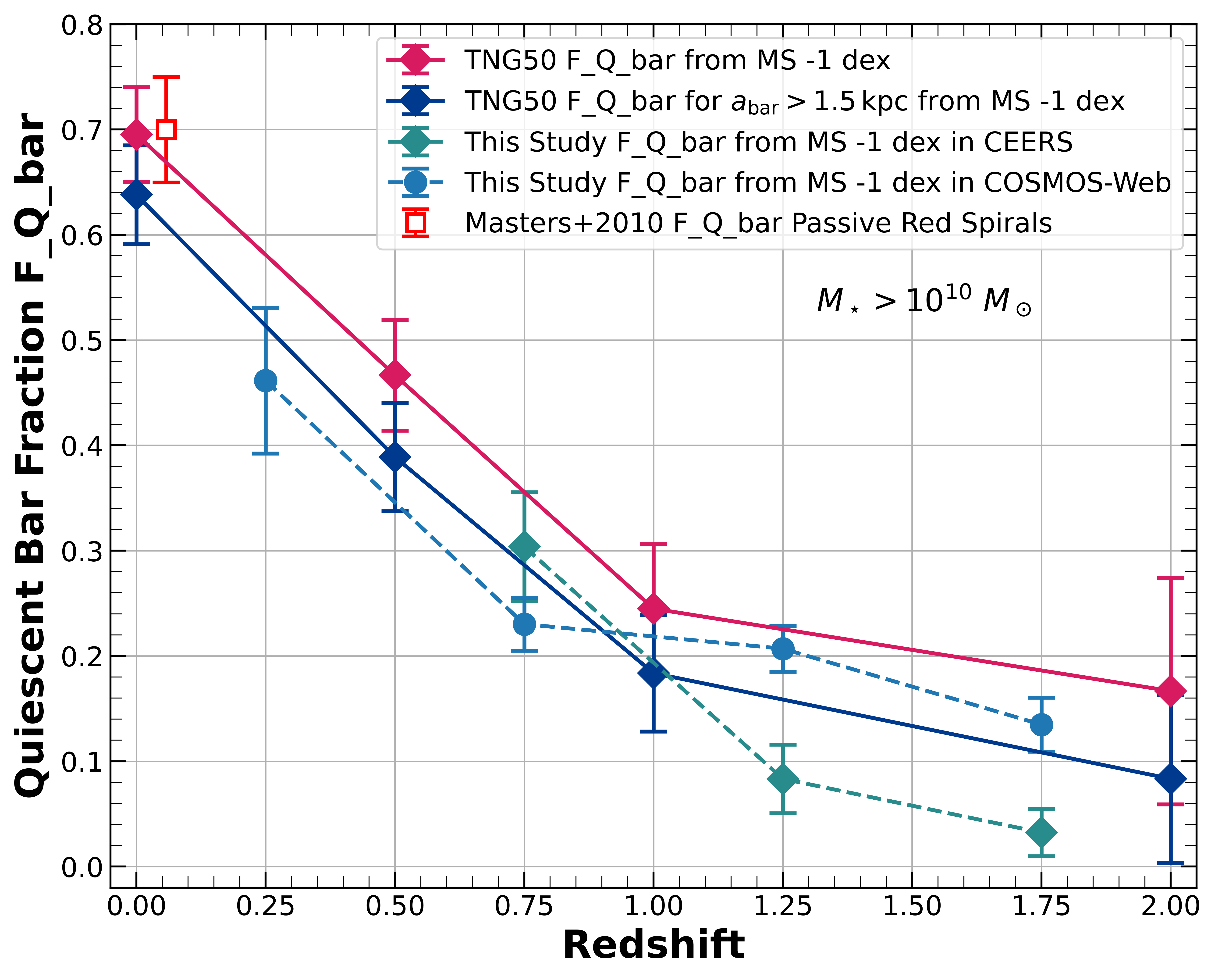}
    \end{subfigure}
    \caption{\textbf{Left}: The quiescent bar fraction in TNG50-1 is plotted for all bar sizes (solid magenta line) and for $a_{\rm bar} > 1.5$ kpc (solid navy blue line) using the sSFR selection method. Also plotted are the observed quiescent bar fractions in CEERS (dashed purple line) and COSMOS-Web (dashed orange line) using the sSFR selection method. The TNG50-1 results for $a_{\rm bar} > 1.5$ kpc agree well with the observed bar fraction in quiescent disk galaxies, rising from $\sim$20.8\% $\pm$ 8.2\% at $z \sim$ 1 to $\sim$65.2\% $\pm$ 4.0\% at $z \sim$~0. \textbf{Right}: Same as left panel, but using the MS -1 dex method to identify quiescent galaxies in both the data and TNG50-1.}
    \label{fig:TNG50_fqbar_ssfr}
\end{figure*}

\subsubsection{Comparison with the Quiescent and Actively Star-Forming Bar Fractions in Simulations
}\label{sec:theory}

In this section, we compare our observational results on the bar fraction in quiescent disk galaxies ($F_{\mathrm{Q\_bar}}$) and in actively star-forming galaxies ($F_{\mathrm{SF\_bar}}$) to bar fractions derived using stellar bars identified in the IllustrisTNG TNG50-1 cosmological simulation \citep{Marinacci-etal-2018,Naiman-etal-2018,Nelson-etal-2018, Springel-etal-2018,Pillepich-etal-2018,Pillepich-etal-2019, Nelson-etal-2019,Nelson-etal-2019-Release}. TNG50-1 follows galaxy formation within a periodic volume of $(51.7$ ${\rm Mpc})^{3}$ using the moving-mesh magnetohydrodynamics code AREPO, which combines adaptive spatial resolution with accurate modeling of gas dynamics \citep{Springel-2010}. The simulation achieves a baryonic mass resolution of $\sim8.5\times10^{4}M_\odot$, a typical spatial resolution of $\sim100$ pc in dense star-forming regions for the gas component, and a typical spatial resolution for the stellar component of $\sim300$ pc, enabling the formation and evolution of resolved stellar disks and bars across cosmic time \citep{Pillepich-etal-2019}.

\begin{figure}[!htb]
    \centering
    \includegraphics[width=0.9\linewidth]{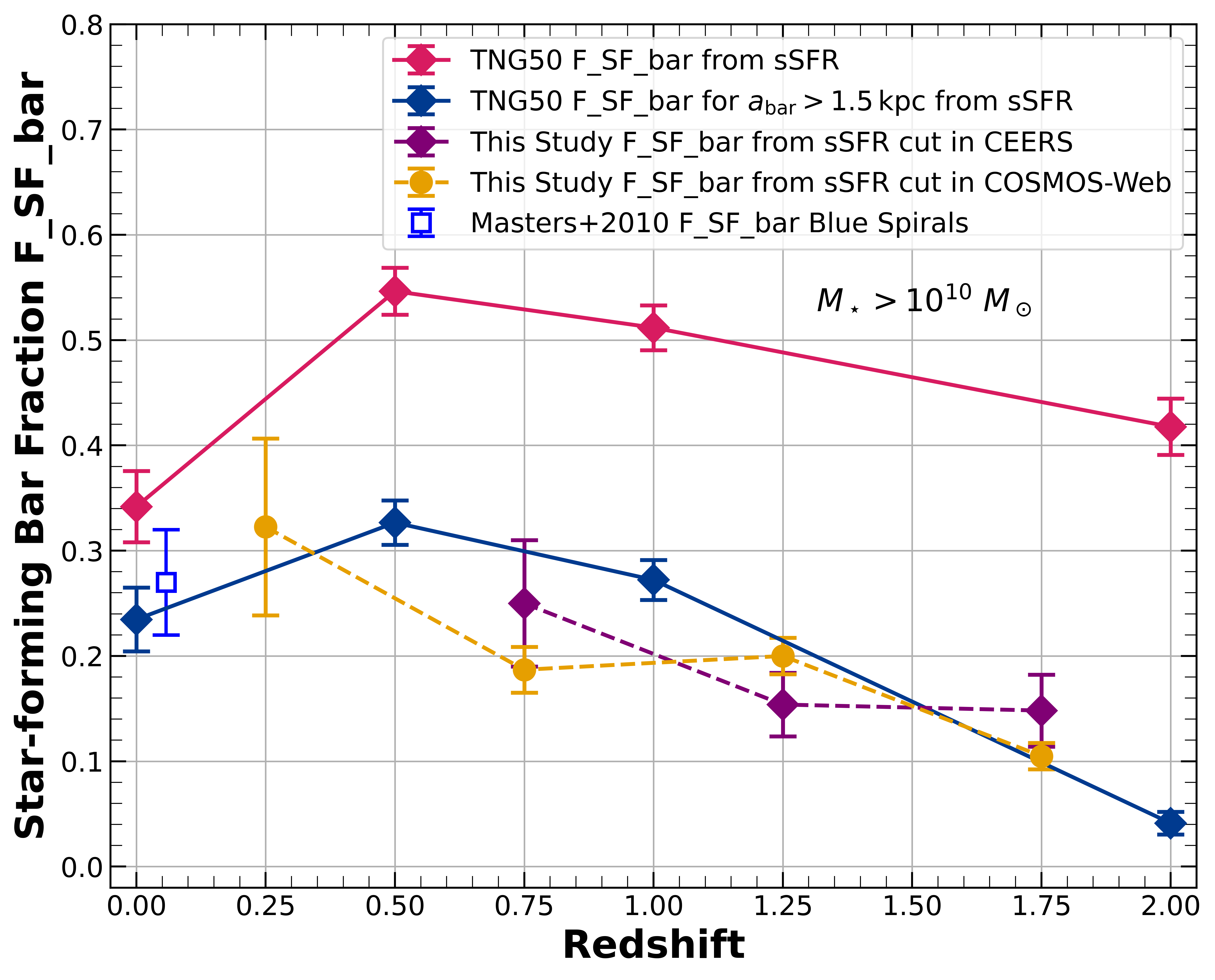}
    \caption{The actively star-forming bar fraction in TNG50-1 is plotted for all bar sizes (solid magenta line) and for $a_{\rm bar} > 1.5$ kpc (solid navy blue line) using the sSFR selection method (Section~\ref{sec:starform id}). Also plotted are the observed actively star-forming bar fractions in CEERS (dashed purple line) and COSMOS-Web (dashed orange line) using the sSFR selection method. The TNG50-1 results for $a_{\rm bar} > 1.5$ kpc agree well with the observed bar fraction in actively star-forming galaxies, rising from $\sim$4.1\% $\pm$ 1.1\% at $z \sim$ 2 to $\sim$32.7\% $\pm$ 2.1\% at $z \sim$ 0.5, and decreasing to $\sim$23.5\% $\pm$ 3.0\% at $z \sim$ 0.}
    \label{fig:TNG50_fsfbar_ssfr}
\end{figure}

 \begin{figure*}[!htb]
    \centering
    \includegraphics[width=0.85\textwidth]{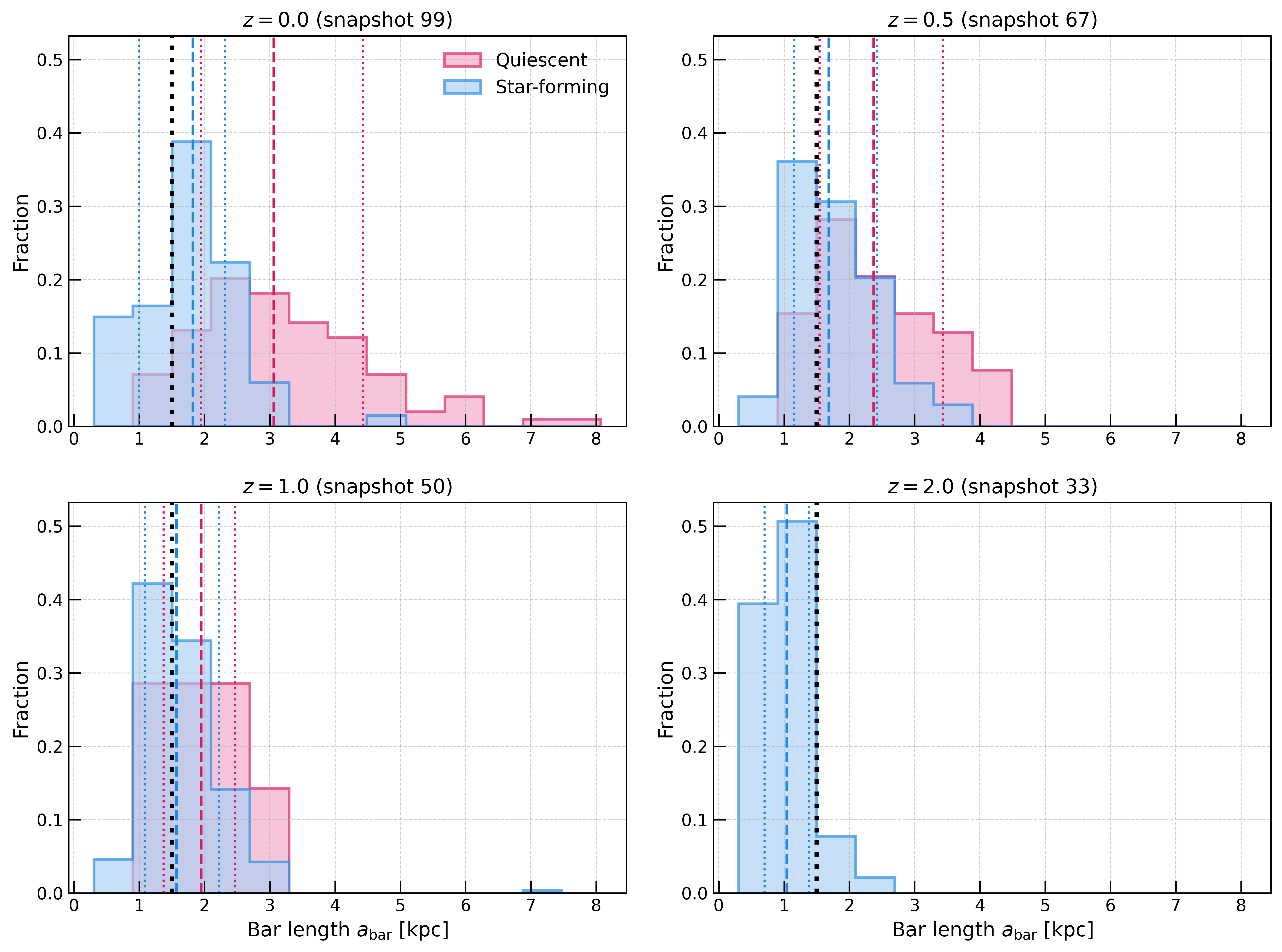}
    \caption{The distributions of bar size for quiescent (red) and actively star-forming (blue) barred galaxies identified via the sSFR method in TNG50-1 are plotted from $z \sim 0$ to $z \sim 2$. The medians of each distribution are plotted as dashed lines in their respective colors with the 16th and 84th percentiles shown as dotted lines. The black dotted lines mark $a_{\mathrm{bar}}$ $=$ 1.5 kpc. From $z \sim 1$ to $z \sim 0$, the median $a_{\mathrm{bar}}$ for quiescent disk galaxies is longer than that of actively star-forming galaxies, and this difference increases from $z \sim 1$ to $z \sim 0$. The distributions of star-forming galaxy bar sizes are skewed toward lower bar sizes with a significant fraction of galaxies with $a_{\mathrm{bar}}$ $<$ 1.5 kpc, while the distributions of quiescent galaxy bar sizes show that quiescent disk galaxies preferentially host bars of larger sizes compared to actively star-forming galaxies.}
    \label{fig:TNG50_barlen}
\end{figure*}  

In the following analysis, we use the catalog presented in \citet{Zana2022}, and supplement it with the SFRs presented in \citet{Donnari2019} and \citet{Pillepich-etal-2019}, which we detail below. Following disk galaxy definitions from \citet{Zana2022}, we adopt the criterion to identify disk galaxies as systems in which the combined stellar mass of the rotating components (thin disk, thick disk, and pseudobulge) accounts for more than 50\% of the total stellar mass ($M_{\rm thin} + M_{\rm thick} + M_{\rm pseudo\mbox{-}bulge} > 0.5 M_\star$), ensuring that rotationally supported structures dominate the stellar mass distribution. 

We identify barred galaxies among the disk galaxies in TNG50-1 by using  the bar classifications presented in  \citet{Zana2022}. Bars were identified using Fourier decomposition of the face-on stellar mass surface density, requiring a prominent $m = 2$ mode and a nearly constant phase over a significant radial extent, ensuring robust identification of coherent, large-scale stellar bars rather than spiral structure \citep{Zana2022}. Stellar masses are taken from the \citet{Zana2022} catalog which defines stellar mass as the sum of the mass of all stellar particles assigned by \textsc{subfind} to the galaxy. SFRs are taken from the time-averaged SFR catalog of \citet{Donnari2019} (first presented in \citealt{Pillepich-etal-2019}), derived from stellar particles formed over fixed time intervals and thus directly comparable to observational tracers. We use SFRs measured within a 30 physical kpc aperture and averaged over a timescale of 10 Myr. We then use these data in order to adopt the previously defined sSFR and main sequence methodologies to identify quiescent and actively star-forming galaxies.

We examine the quiescent and actively star-forming bar fractions of the full sample of disk galaxies with stellar masses $M_\star > 10^{10} M_\odot$ via two separate measures: with bars of all sizes and with a restriction of bar semi-major axis $a_{\rm bar} > 1.5$ kpc as ellipse fits can robustly trace bars with projected semi-major axes $a_{\mathrm{bar}}$ $>$ 1.5 kpc ($\sim$ 2 $\times$ the PSF in F200W images). We use the measure $R_\Phi$ from \citet{Zana2022} to restrict bar size, where $R_\Phi$ is defined as the outer radius at which the phase of the $m = 2$ Fourier mode remains approximately constant relative to its value at the radius where bar strength peaks, and is adopted as the bar semi-major axis.

In Figure~\ref{fig:TNG50_fqbar_ssfr}, we present the quiescent bar fraction ($F_{\mathrm{Q\_bar}}$) in TNG50-1 for all bar sizes and for $a_{\mathrm{bar}}$ $>$ 1.5 kpc, using the sSFR methodology (Left Panel) and main sequence methodology (Right Panel) to identify quiescent galaxies, as outlined in  Section~\ref{sec:quiescent id}. It is very striking to see that, in TNG50-1, when we restrict the analysis to larger bars with $a_{\mathrm{bar}}$ $>$ 1.5 kpc, the quiescent bar fraction only decreases slightly (by less than 10\%) compared to the total bar fraction at all redshifts.  This has the important implication that {\textit{quiescent barred galaxies in TNG50-1 tend to host long, well-developed bars}}.  One possible interpretation of this trend is that as barred galaxies evolve and get stronger (longer), they also evolve toward quiescence. We will discuss this more in Section~\ref{sec:discussion}.

We also compare the fraction of actively star-forming galaxies ($F_{\mathrm{SF\_bar}}$) that are barred in TNG50-1 for all bar sizes and for $a_{\mathrm{bar}}$ $>$ 1.5 kpc in Figure~\ref{fig:TNG50_fsfbar_ssfr}, using sSFR to identify actively star-forming galaxies. In this case we see that when we restrict the analysis to larger bars with $a_{\mathrm{bar}}$ $>$ 1.5 kpc, the bar fraction among actively star-forming galaxies falls significantly compared to the total bar fraction for all bar sizes. This large decrease implies that {\textit{actively star-forming barred galaxies in TNG50-1 host a large population of small bars (with $a_{\mathrm{bar}}$ $<$ 1.5 kpc) from $z \sim$ 2 to $z \sim$ 0.5.}}

Following from our analysis of bar fractions in TNG50-1 for all bar sizes and for $a_{\mathrm{bar}}$ $>$ 1.5 kpc in quiescent and actively star-forming galaxies, we now compare the bar sizes of quiescent (identified via the sSFR method) and actively star-forming (identified via the sSFR method) barred galaxies in TNG50-1. Figure~\ref{fig:TNG50_barlen} shows distributions of bar size for quiescent (red) and actively star-forming (blue) barred galaxies in TNG50-1 from $z \sim 0$ to $z \sim 2$. The medians of each distribution are plotted as dashed lines in their respective colors, and the black dotted lines mark $a_{\mathrm{bar}}$ $=$ 1.5 kpc. From $z \sim 1$ to $z \sim 0$, the median $a_{\mathrm{bar}}$ for quiescent barred galaxies is longer than that of actively star-forming galaxies, and the difference between the median bar length of quiescent and actively star-forming galaxies increases from $z \sim 1$ to $z \sim 0$. The distributions of star-forming galaxy bar sizes are skewed toward lower bar sizes with a significant fraction of galaxies with $a_{\mathrm{bar}}$ $<$ 1.5 kpc, while the distributions of quiescent galaxy bar sizes show that quiescent barred galaxies preferentially host bars of larger sizes compared to actively star-forming galaxies.

\section{Discussion}\label{sec:discussion}

In this section we summarize our main results and discuss their implications.

The role of stellar bars on SF, quenching, and galaxy evolution has been studied in the nearby Universe, but this topic remains largely unexplored at higher redshifts. In this paper, we performed one of the first systematic studies of barred galaxies from $z\sim0$ to $z\sim2$, examining the relationship between bars, SF activity, and host galaxy properties, and reported three key findings:

\begin{enumerate}
\item At high redshift ($z \sim 1$--$2$)  barred galaxies tend to have predominantly high sSFRs and low Sérsic indices ($n \leq 2$), while at low redshifts we see the emergence of barred galaxies that have both low sSFR and higher $n$ ($ > 2$) (Section~\ref{sec:host properties}; Figures~\ref{fig:n_ssfr} and~\ref{fig:n_ssfr_cos}), corresponding to higher quiescence and higher central light concentrations.
 
\item From $z\sim2$ to $z\sim0$, the fractional contribution of barred quiescent disk galaxies to the bar fraction rises significantly while the fractional contribution of barred actively star-forming galaxies to the bar fraction falls over time from $z\sim2$ to $z\sim0$ (Section~\ref{sec:SF Q contribution}; Figure~\ref{fig:contribute_CEERS}).   

\item Over the last 10 Gyr (from $z \sim2$ to $z \sim0$), the fraction of quiescent disk galaxies that are barred ($F_{\mathrm{Q\_bar}}$) rises steeply from $\sim2.3\%\pm2.3\%$ at  $z \sim 1.75$ to $\sim54.5\%\pm7.5\%$ at $z \sim 0.25$, while the fraction of actively star-forming disk galaxies that are barred ($F_{\mathrm{SF\_bar}}$) rises at a shallower rate (Section~\ref{sec:quiescent SF fbar CEERS COSMOS}; Figure~\ref{fig:F_Q_bar}). While $F_{\mathrm{SF\_bar}}$ is higher than the quiescent bar fraction ($F_{\mathrm{Q\_bar}}$) at high redshift ($z\sim2$), it is surpassed by a higher $F_{\mathrm{Q\_bar}}$ by $z\sim0.25$. 
\end{enumerate}

Taken together, the above results suggest that over the last 10 Gyr barred galaxies are changing over time, with \textit{many becoming galaxies with lower sSFR (higher quiescence) and higher central light concentrations}.

A full interpretation of these results is complex both due to observational and theoretical
factors. From an observational perspective, our study has both strengths and limitations. A major strength of our study is that our results are robust and not overly sensitive to the methods for classifying bars or identifying quiescent galaxies, the survey details, and the sample selection. In particular, we have shown throughout this paper that we obtain similar results using two independent bar-classification methods (ellipse fits in CEERS and neural networks trained on visual classifications in COSMOS-Web) and two methods to identify quiescent galaxies (based on sSFR and the main sequence; Section~\ref{sec:quiescent id}). Similarly, the agreement between analyses performed on two galaxy samples based on two widely different surveys (the deep, small-area CEERS survey and the shallower, large-area COSMOS-Web survey) suggests our results are fairly robust against survey specifications, sample biases, and cosmic variance. At the same time, several limitations should be noted. Our adoption of a fixed and relatively narrow stellar mass range across all redshift bins may complicate evolutionary interpretations, as galaxies occupying the same mass range at different epochs are not progenitors or descendants of one another. Additionally, our study can only detect bars with projected semi-major axis lengths $a_{\mathrm{bar}}$ $>$ 1.5 kpc, which can be resolved and robustly detected by ellipse fits. As such, we may be missing a population of small bars, especially at higher redshifts when galaxies themselves are smaller. Consequently, the results reported here should not be interpreted as applying to the true population bars, but rather those we are able to detect.

From a theoretical perspective, the formation, lifetime, evolution, and impact of bars is an area of active research (see Sections~\ref{sec:scen1} and \ref{sec:scen2}). There are multiple formation pathways for bars including spontaneous and tidally triggered local and global instabilities. At higher redshifts where physical conditions in galaxies are different (e.g., stellar disk mass, disk-to-halo mass fraction, gas fractions, turbulence), the formation and evolution of barred galaxies may be significantly different from what we know at low redshifts. Additionally, the presence of a bar at a given time does not necessarily imply that the bar has always been present. Bars may have formed recently, dissolved and reformed, or weakened over cosmic time.

Notwithstanding the complexity of this landscape, we focus on three specific questions around bars and quiescence in Sections~\ref{sec:scen1}, \ref{sec:scen2}, and \ref{sec:scen3} below. In Section~\ref{sec:scen1} we discuss if and how bars can contribute to quiescence, in Section~\ref{sec:scen2} we discuss if bars are more likely to form and survive in quiescent and gas-poor galaxies, and in Section~\ref{sec:scen3} we discuss broader considerations on barred galaxy evolution.

\subsection{Do bars contribute to quiescence?}\label{sec:scen1}

\begin{figure*}
    \centering
    \includegraphics[width=0.99\textwidth]{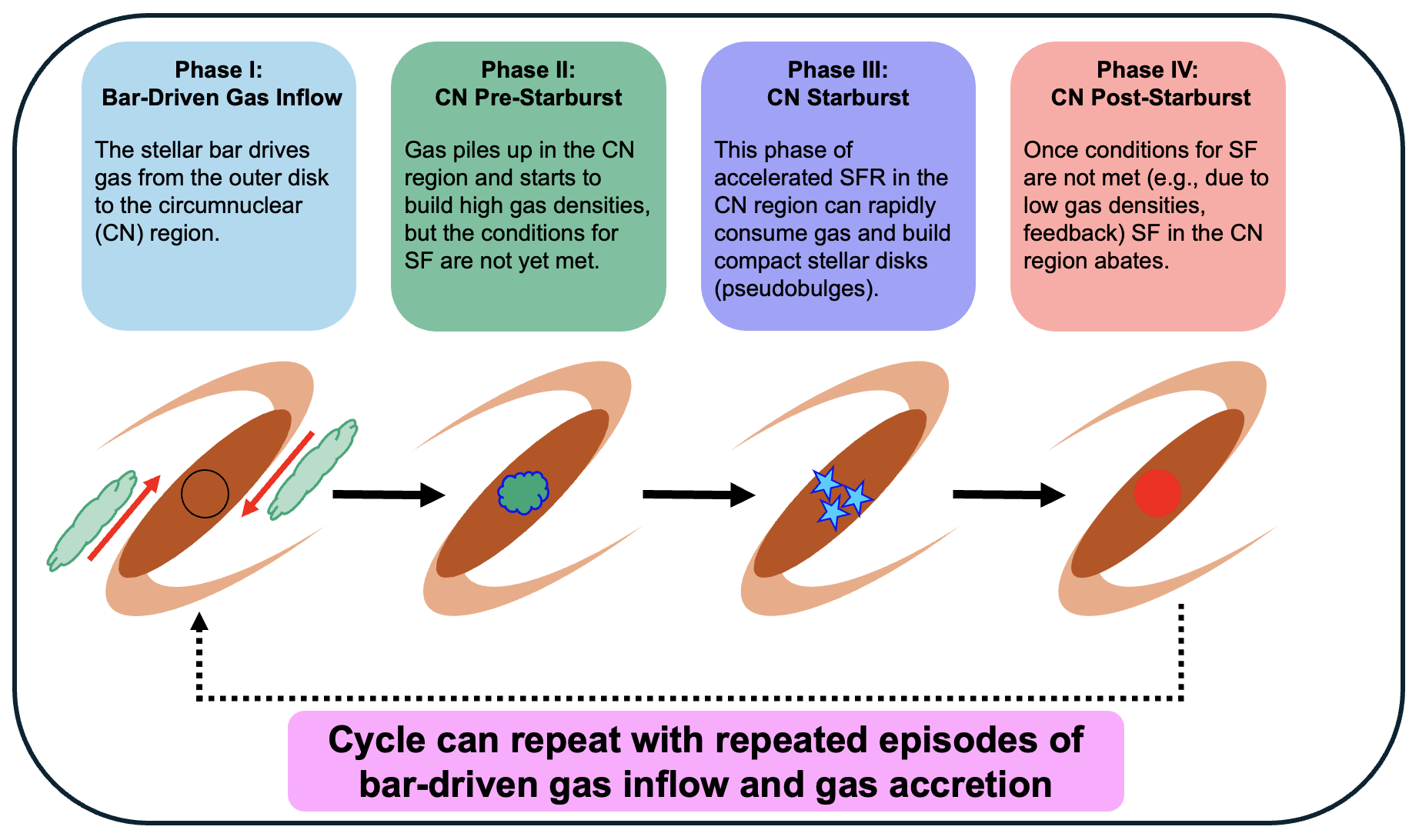}
    \caption{Schematic representation of the phases of bar-driven evolution described in Section~\ref{sec:scen1}. A barred galaxy can evolve from a bar-driven gas inflow phase to the CN pre-starburst, CN starburst, and CN post-starburst phases. The cycle can repeat as long as the galaxy is accreting gas. Over time, the repeated effects of bar-driven gas inflow and accelerated SF activity in the CN region can help make a galaxy quiescent by redistributing gas from the outer disk into the CN region and accelerating the conversion of this gas into stars. At low redshifts, once the barred galaxy stops accreting gas, the galaxy can stay in this relatively quiescent post-starburst phase.} 
    \label{fig:Main_Figure}
\end{figure*}

One possible interpretation of our findings is that over the last 10 Gyr, bars are leading to quiescence. One pathway for this to happen is illustrated in Figure~\ref{fig:Main_Figure}, which shows the four phases of evolution of a barred galaxy: 
 
\begin{itemize}
\item \textit{Phase I - Bar-Driven Gas Inflow}: The bar drives gas located in the outer disk inside the corotation resonance of the bar into the CN region. Typically, the SF efficiency (SFE) along strong bars is low due to shear and non-circular motions and consequently, only a small fraction of the gas is converted into stars along the bar and most of it reaches the CN region. This phase has been observed at $z\sim0$ in barred galaxies (e.g., \citealt{Regan1997, Jogee2005,George2020,Geron2024}).
 
\item \textit{Phase II - Circumnuclear Pre-Starburst}: In this phase, gas piles up in the CN region and starts to build high gas densities, but the conditions for SF are not yet met.
 
\item \textit{Phase III - Circumnuclear Starburst and Accelerated SF}: Once the conditions for SF are met (e.g., high gas densities, low gas velocity dispersion, low shear), high SFRs are triggered in the CN region. This phase of accelerated SF can build compact CN disks known as disky bulges or pseudobulges. This phase has been observed in the CN region of local barred galaxies where large gas densities (several 1000 $M_\odot$ pc$^{-2}$) and SFR densities ($>$ 10$^{-6}$ $M_\odot$ pc$^{-2}$ yr$^{-1}$) are measured (e.g., \citealt{Jogee2005}) and pseudobulges are common (e.g., \citealt{KormendyKennicutt2004}). 
 
\item \textit{Phase IV - Circumnuclear Post-Starburst}: Once conditions for SF are no longer met (e.g., due to feedback, low gas densities, etc.) SF in the CN region can abate.

\end{itemize}

As long as the barred galaxy is accreting gas, the above four phases can repeat, with the galaxy moving from a bar-driven gas inflow phase to a CN starburst phase to a CN post-starburst phase. Over time, the repeated effects of bar-driven gas inflow and the resulting accelerated SF activity in the CN region can help make a galaxy quiescent by effectively redistributing the gas from the outer disk into the CN region and accelerating the conversion of this gas into stars.  At low redshifts, once the barred galaxy stops accreting gas, the galaxy can stay in this relatively quiescent post-starburst phase.

\begin{figure}
    \centering
    \includegraphics[width=0.99\linewidth]{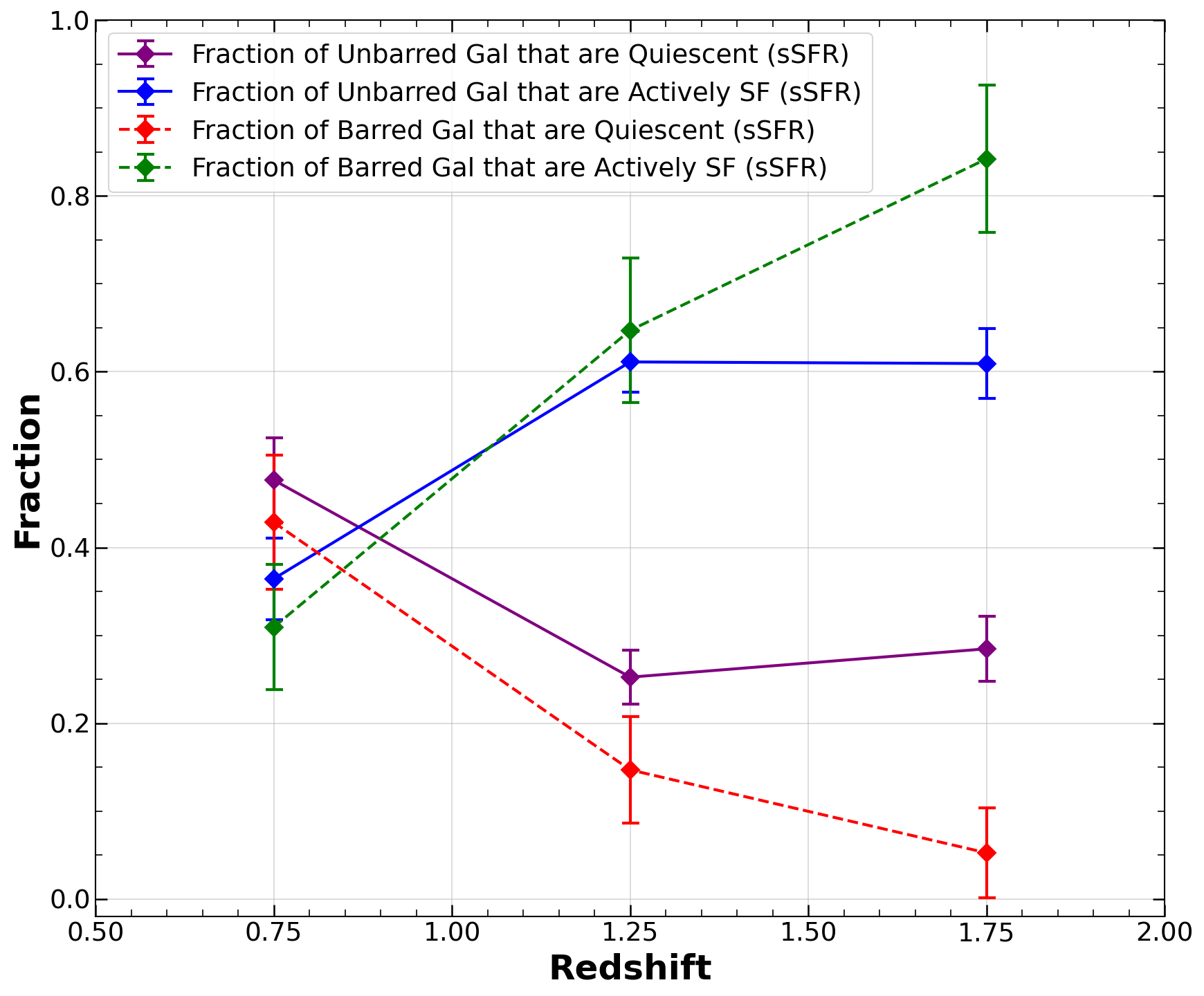}
    \caption{The fraction of barred (red dashed line) and unbarred (purple solid line) disk galaxies that are quiescent, along with the fraction of barred (green dashed line) and unbarred (blue solid line) disk galaxies that are actively star-forming, as a function of redshift.}
    \label{fig:bar_unbar}
\end{figure}

However, as we discuss in more detail in Section~\ref{sec:scen3}, there are mechanisms other than bars that can induce
 quiescence in galaxies. For example, tidal interactions can lead to large gas inflows, accelerated SF, bulge-building, and eventually quiescence (e.g., \citealt{Barnes1992} and references therein; \citealt{Gnedin2003}). Quiescence can also be induced by a decline in the galaxy’s gas supply from cold mode accretion at high halo masses, and by environmental effects in galaxy clusters, such as strangulation, starvation, and ram pressure stripping (e.g., \citealt{Gunn1972, Larson1980, Koopmann2004,Crowl2005,Singh2019, Gentile2025}). Therefore, while bars play a role in inducing quiescence, they may not be the sole or dominant mechanism by which galaxies become quiescent. Indeed, in Figure~\ref{fig:n_ssfr} and Figure~\ref{fig:n_ssfr_cos}, unbarred disks occupy the low sSFR and high Sérsic index regimes even at high redshift, lending more support for the notion that there are other pathways by which galaxies become quiescent and build higher central light concentrations. In Figure~\ref{fig:bar_unbar} we plot the quiescent fractions of both barred (dashed red line) and unbarred (solid purple line) disk galaxies, and observe that they both rise from $z\sim2$ to $z\sim0.5$. As both the barred and unbarred disk populations are becoming increasingly quiescent, there is no compelling evidence that bars are doing all of the work to induce quiescence.

\subsection{Are bars more likely to form and survive in quiescent and gas-poor galaxies?}\label{sec:scen2}

Bars may form spontaneously through the growth of local and global non-axisymmetric instabilities in sufficiently massive, dynamically cold disks (e.g., \citealt{Toomre1981,Ostriker1973,Athanassoula2003}; D'Onghia et al. 2026, in prep.), or they may be triggered by tidal interactions and galaxy encounters \citep[e.g.,][]{Noguchi1987,Miwa1998,Bi2022,Rosas2024}. The formation, lifetime, and impact of a bar depend on several coupled factors, including the structure of the stellar disk, the properties of the dark matter halo, the relative importance of ordered and random motions, and the abundance and turbulent state of the gas. An additional complication is that the presence of a bar at a given epoch does not imply that the bar has been present for the entire lifetime of the galaxy; bars may form, grow, weaken, be destroyed, or reform over time. Recently, \citet{Hawthorn2024} found that, above a limiting value, the bar-formation timescale decreases exponentially with increasing baryonic disk mass fraction relative to the dark matter halo. After formation, bars can grow stronger, longer, and slower by losing angular momentum to a live dark matter halo \citep{Athanassoula2003, DOnghia2020}. This angular momentum exchange causes the corotation resonance to move outward, allowing the bar to lengthen and exert stronger gravitational torques. Conversely, bars can be weakened or destroyed by large central mass concentrations (CMCs), which can induce chaotic orbits or shift material onto $x_2$ orbits rather than the bar-supporting $x_1$ orbital family \citep{Shen2004,Athanassoula2005cmc}. \citet{Athanassoula2005cmc} argue that present-day galaxies are robust against destruction by CMCs as they do not host large enough CMCs, even if one considers supermassive black holes, molecular gas disks, disky bulges, and other central structures. However, at higher redshifts, where galaxies are more gas-rich and host smaller disks, bars may be more susceptible to destruction by CMCs.

The impact of gas on bar formation and destruction is complex. On the one hand, gas-rich disks can be dynamically cold and susceptible to the growth of large-scale $m=2$ modes, potentially facilitating bar formation in massive high-redshift disks \citep[e.g.,][]{Bournaud2002,Romano2008,Kraljic2012,Spinoso2017,Rosas2022}. On the other hand, several studies suggest a high gas fraction can cause bars to weaken or to experience slower growth. Gas inflows can build central mass concentrations that weaken or destroy some bars \citep[e.g.,][]{Shen2004,Bournaud2005,Athanassoula2005cmc,Debattista2006}, while massive gas clumps can sink by dynamical friction and dynamically heat the stellar disk \citep{Shlosman1993}. Gas-rich disks may also reduce bar growth by limiting the angular momentum exchange that allows bars to slow down and lengthen. For example, \citet{Athanassoula2013} showed that increasing the gas fraction delays bar formation, slows subsequent bar growth, and produces weaker bars by the end of the simulations. Similarly, recent simulations suggest that high gas fractions can keep bars faster and shorter, reducing their ability to grow through secular angular momentum exchange \citep{Beane2023}. 

In summary, these different studies suggest that bars are likely to form in gas-rich massive dynamically cold disks, but once the bar forms, a high gas fraction can slow down bar growth or lead to CMCs that may weaken or destroy the bar. Within this framework, the observed increase with time of the quiescent bar fraction may represent an evolutionary sequence where massive gas-rich disks at high redshift first develop shorter and dynamically younger bars embedded within rapidly evolving disks. As galaxies evolve to lower redshifts, repeated bar-driven gas inflows lead to rapid gas consumption via CN starbursts, the formation of disky bulges, and declining gas fractions (Figure~\ref{fig:Main_Figure}), which allow the bar to strengthen as the galaxy transitions toward quiescence. Our earlier finding that in TNG50-1 quiescent barred galaxies tend to host long, well-developed bars, while star-forming barred galaxies tend to host a large population of small bars (Section~\ref{sec:theory}) supports this evolutionary scenario. In this picture, \textit{the increasing association between bars and quiescent galaxies at lower redshift reflects both bar-driven secular evolution and the evolving dynamical state of galaxy disks over cosmic time.}

 \subsection{Broader considerations on barred galaxy evolution}\label{sec:scen3}

 The interpretation of our results must also be considered within the broader context of galaxy evolution, where the evolution and impact of a bar depends on gas accretion, environment, feedback processes, and the lifetime of the bar itself.

The availability of cold gas is expected to play a key role in shaping galaxy evolution. Cosmological simulations indicate that cold gas accretion from the intergalactic medium is a dominant source of SF fuel at $z$ $>$ 1 \citep[e.g.,][]{Katz2003, Keres2005, Brooks2009, Dekel2009}. In halos below the critical mass threshold for virial shock heating, gas is efficiently delivered via cold, filamentary streams \citep[e.g.,][]{Dekel2006}, sustaining the high gas fractions that promote strong bar-driven inflows toward the CN regions. At higher halo masses, where hot-mode accretion and long cooling times prevail, the reduced availability of cold gas limits a bar’s ability to sustain recurrent SF activity. Nevertheless, secondary gas accretion mechanisms, such as gas acquired through minor mergers, can intermittently replenish the gas reservoir \citep[e.g.,][]{Sancisi2008, Angles2017}, potentially reactivating the bar-SF cycle even in more massive or evolved systems.

The evolution of barred galaxies may also be influenced by the history of tidal interactions and mergers. Bars may form spontaneously through the growth of local and global instabilities (e.g., \citealt{Toomre1981,Ostriker1973,Athanassoula2003}; D'Onghia et al. 2026, in prep.), or they may be triggered by tidal interactions and galaxy encounters \citep[e.g.,][]{Noguchi1987,Miwa1998,Bi2022,Rosas2024}. Using the TNG50 simulation, \citet{Rosas2024} investigated the role of mergers and galaxy interactions in the formation and destruction of bars. They found that interactions can both trigger the formation of bars and contribute to their dissolution, depending on the properties of the interaction and the host galaxy. Their results further suggest that while many bars survive for extended periods of time, tidal interactions and mergers can significantly alter the evolutionary pathways of barred galaxies by modifying the structure and dynamics of their host disks.

The evolution of barred galaxies also depends on their large-scale environment, specifically on whether they are in field, group, or cluster environments.  Environmental effects in galaxy clusters, such as strangulation, starvation, and ram pressure stripping (e.g., \citealt{Gunn1972, Larson1980, Koopmann2004,Crowl2005,Singh2019, Gentile2025}) can reduce the gas supply of barred galaxies. As outlined in Section~\ref{sec:scen2}, the gas fraction plays an important role in the evolution of bars, and high gas fractions can slow down bar growth or lead to CMCs that may weaken or destroy the bar.  

Feedback processes may provide an additional source of regulation. Cosmological simulations indicate that supernova feedback can delay, and in some cases suppress, bar formation by altering the assembly and dynamical state of the stellar disk, whereas black hole feedback appears to have a negligible impact on whether a bar forms \citep{Rosas2025}. Bars are not expected to directly fuel AGN in present-day barred galaxies as bar-driven gas inflows tend to stall in the inner few hundred pc near the Inner Lindblad Resonances (ILRs) and additional transport mechanisms are needed to drive the gas down to the central black hole (\citealt{Buta1996}; review by \citealt{Jogee2006} and references therein). However, some studies do find that AGN host galaxies are marginally more likely to host a bar (e.g., \citealt{Galloway2015,Garland2023}) and that barred active galaxies experience excess nuclear activity in comparison to unbarred active galaxies (e.g., \citealt{Alonso2013,Marels2025}). Simulations further show that bar formation can be accompanied by enhanced central SF and black hole accretion, potentially contributing to the depletion of the central gas reservoir and the subsequent reduction of SF in the nuclear regions \citep{Rosas2020}.

Finally, the timescales over which bars influence their host galaxies remain an active area of research. The TIMER (Time Inference with MUSE in Extragalactic Rings) survey has provided important observational constraints on bar-driven secular evolution by studying the stellar populations and kinematics of nearby barred galaxies with MUSE integral-field spectroscopy \citep{Gadotti2019,Gadotti2020,Bittner2020}. TIMER observations show that nuclear discs and rings assembled through bar-driven evolution are often several Gyr old, implying that bars can survive for long periods of time. More recently, \citet{deSaFreitas2025} derived bar ages for 20 nearby galaxies and found formation epochs spanning between $1$ and $13$ Gyr ago. They further found that galaxies hosting older bars tend to be more quenched, consistent with a picture in which long-lived bars contribute to the gradual redistribution and depletion of cold gas over cosmic time. Together, these results suggest that the evolutionary impact of bars depends not only on their presence, but also on the availability of gas and the length of time over which secular processes are able to operate.
 
\section{Summary}\label{sec:summary}

Stellar bars are considered to be powerful drivers of secular evolution in disk galaxies (e.g., \citealt{KormendyKennicutt2004} and references therein, \citealt{Athanassoula2003, Jogee2005}). Bars exert gravitational torques and shocks which efficiently channel gas from the outer disk into the CN region, 
establishing several bar-driven evolutionary phases of SF (see Figure~\ref{fig:Main_Figure}).

Barred galaxies and their relationship with SF activity, the onset of quiescence, and host galaxy properties have been studied at $z\sim0$ (e.g., \citealt{Jogee2005,Masters2010b, Masters2010a, Masters2012,Cheung2013, Ellison2011, Khoperskov2018, Lin2020, Fraser2020, Newnham2020, George2020, George2021}), with some studies extending to intermediate redshifts (e.g., \citealt{Cameron2010,Melvin2014}), but higher redshifts remain unexplored.

We present one of the first studies of the relationship between bars, SF activity, and host galaxy properties over the last 10 Gyr from $z\sim2$ to $z\sim0$. We use a mass-complete sample of 1,171 \textit{JWST} CEERS galaxies with $M_\star > 10^{10}\ M_\odot$ and bar classifications based on ellipse fits from \citet{Guo2025} at $z\sim0.5$ to $z\sim2$. We also cross-check our analysis using a sample of 14,117 galaxies at $z\sim0$ to $z\sim2$ from COSMOS-Web, where bars and disks are identified from neural network classifications \citep{HuertasCompany2025}. We identify quiescent galaxies using two different methods: (a) sSFR $< 10^{-11}\ {\rm yr}^{-1}$ and (b) a location $> 1$ dex below the star-forming main sequence on the stellar mass-SFR plane. We identify actively star-forming galaxies as those with sSFR $> 10^{-10}\ {\rm yr}^{-1}$. 

A major strength of our study is that our results are robust and not overly sensitive to the methods for classifying bars or identifying quiescent galaxies, the survey details, the sample selection, and cosmic variance as we use two different bar-classification methods (ellipse fits in CEERS and neural networks trained on visual classifications in COSMOS-Web), two methods to identify quiescent galaxies, and two different surveys (CEERS
and COSMOS-Web). We find several new key results on barred galaxies and report trends as a function of redshift, which, to the best of our knowledge, have not been unveiled before:

\begin{enumerate}

\item At high redshift ($z\sim1$--$2$), barred galaxies tend to have predominantly high sSFRs, while at low redshifts a large number of barred galaxies occupy the low sSFR regime. Additionally, at high redshift ($z\sim1$--$2$), barred galaxies predominantly have Sérsic indices $n\leq2$, but at low redshifts, \textit{we see the emergence of barred galaxies which have both low sSFR and higher Sérsic indices ($n > 2$), suggestive of quiescent systems with central bulges} (Figures~\ref{fig:n_ssfr} and~\ref{fig:n_ssfr_cos}).

\item \textit{The fractional contribution of barred quiescent disk galaxies to the bar fraction rises significantly from $z\sim2$ to $z\sim0$} (Figure~\ref{fig:contribute_CEERS}). With the sSFR method of identifying quiescent galaxies, in CEERS barred quiescent disk galaxies contribute to less than $\sim8\%$ (1\%/12\%) of the bar fraction at $z\sim1.75$ and as much as $\sim43\%$ (12\%/28\%) by $z\sim0.75$. In COSMOS-Web, barred quiescent disk galaxies contribute $\sim4\%$ (0.5\%/12\%) to the bar fraction at $z\sim1.75$ and as much as $\sim33\%$ (13\%/40\%) by $z\sim0.25$. \textit{In contrast, the fractional contribution of barred actively star-forming galaxies to the bar fraction falls over time from $z\sim2$ to $z\sim0$.} In CEERS, barred actively star-forming galaxies contribute $\sim75\%$ (9\%/12\%) of the bar fraction at $z\sim1.75$ and only $\sim28\%$ (8\%/28\%) by $z\sim0.75$. In COSMOS-Web, barred actively star-forming galaxies contribute $\sim67\%$ (8\%/12\%) of the bar fraction at $z\sim1.75$ and only $\sim12\%$ (5\%/40\%) by $z\sim0.25$.

\item \textit{The fraction of quiescent disk galaxies that are barred ($F_{\mathrm{Q\_bar}}$) rises steeply over the last 10 Gyr ($z \sim2$ to $z \sim0$) from $\sim2.3\%\pm2.3\%$ at  $z \sim 1.75$ to $\sim54.5\%\pm7.5\%$ at $z \sim 0.25$} (Figure~\ref{fig:F_Q_bar}). We rule out that our results are driven by systematic effects tied to disk sizes (Figure~\ref{fig:large_fbar}). In contrast, the fraction of actively star-forming disks that are barred ($F_{\mathrm{SF\_bar}}$) rises at a much shallower rate, from $\sim10.5\%\pm1.3\%$ at $z \sim 1.75$  to $\sim32.3\%\pm8.4\%$ by $z \sim 0.25$. While $F_{\mathrm{SF\_bar}}$ is higher than the quiescent bar fraction ($F_{\mathrm{Q\_bar}}$) at high redshift ($z\sim2$), it is surpassed by a higher $F_{\mathrm{Q\_bar}}$ by $z\sim0.25$.

\item We explore quiescent and actively star-forming bar fractions in the IllustrisTNG TNG50-1 cosmological simulations: \textbf{a)} We find that restricting bar sizes to $a_{\mathrm{bar}}$ $>$ 1.5 kpc (which can be robustly identified in observations) in TNG50-1 leads to a large drop in the bar fraction in actively star-forming galaxies, especially at higher redshift ($z > 1.5$), but does not significantly change the bar fraction in quiescent disk galaxies (Figures~\ref{fig:TNG50_fqbar_ssfr} and~\ref{fig:TNG50_fsfbar_ssfr}). \textit{This indicates that quiescent barred galaxies in TNG50-1 largely host longer, more well-developed bars than actively star-forming barred galaxies.} \textbf{b)} We also find that after restricting bar sizes to $a_{\mathrm{bar}}$ $>$ 1.5 kpc, TNG50-1 simulations are in good agreement with our observed steep rise from $z\sim$~2 to $z\sim$~0 in the bar fraction of quiescent disk galaxies ($F_{\mathrm{Q\_bar}}$) and the relatively shallow evolution of the bar fraction of actively star-forming galaxies ($F_{\mathrm{SF\_bar}}$).   

\end{enumerate}

Taken together, the above results suggest that over the last $\sim$10 Gyr barred galaxies are changing over time, with \textit{many becoming galaxies with lower sSFR and higher central light concentrations, suggestive of bulges}.
 
Our results allow for the possibility that \textit{bars may lead to quiescence through bar-driven secular evolution}. Specifically, over time the repeated effects of bar-driven gas inflow and the resulting accelerated SF activity in the CN region can help make a galaxy quiescent by effectively redistributing the gas from the outer disk into the CN region and accelerating the conversion of this gas into stars (Figure~\ref{fig:Main_Figure}).  At low redshifts, once the barred galaxy stops accreting gas, the galaxy can stay in this relatively quiescent post-starburst phase. However, quiescence can also be induced by mechanisms other than bars, including tidal interactions, the decline in cold mode accretion, and environmental effects in galaxy clusters, such as strangulation, starvation, and ram pressure stripping. Another possibility allowed by our results is that \textit{bars are more likely to persist and grow in gas-poor, quiescent galaxies}. Lower gas fractions make it less likely to build high CMCs that weaken or destroy the bar, reduce  tidal heating of the stellar disk from gas clumps sinking via dynamical friction,  and reduce the angular momentum gained by the bar, allowing it to grow longer and stronger.
 
Our overall results, and in particular the observed increase of the quiescent bar fraction with time over 10 Gyr, may represent an evolutionary sequence. In this sequence, massive gas-rich disks at high redshift first develop shorter and dynamically younger bars embedded within rapidly evolving disks. As galaxies evolve to lower redshifts, repeated bar-driven gas inflows lead to rapid gas consumption via CN starbursts, the formation of disky bulges, and declining gas fractions (Figure~\ref{fig:Main_Figure}), which allow the bar to strengthen as the galaxy transitions toward quiescence. In this picture, the increasing association between bars and quiescent galaxies at lower redshift reflects both bar-driven secular evolution and the evolving dynamical state of galaxy disks over cosmic time.

\section*{Acknowledgements}
E.W., S.J., and Y.G. acknowledge support from NSF grant AST 2244278, the Rex G. Baker, Jr. and McDonald Centennial Professorship, and the College of Natural Sciences at UT Austin. We acknowledge support from NASA through STScI ERS award JWST-ERS-1345. T.G. is a Canadian Rubin Fellow at the Dunlap Institute. The Dunlap Institute is funded through an endowment established by the David Dunlap family and the University of Toronto.

The authors thank Yetli Rosas-Guevara for sharing information on the TNG50-1 simulations. The authors acknowledge the Texas Advanced Computing Center (TACC) at the University of Texas at Austin for providing HPC and visualization resources that have contributed to the research results reported within this paper. URL: \url{http://www.tacc.utexas.edu}

\newpage

\appendix

In Section~\ref{sec:quiescent id}, we defined a main sequence on the stellar mass-SFR plane in order to identify quiescent galaxies using the main sequence method. We used the published main sequence from \citet{Popesso2023}, which uses a wide set of studies to define the main sequence consistently across a range of redshifts. We utilized the main sequence parameterization presented in Equation 10 and the best-fit parameters reported in Table 2 of \citet{Popesso2023}, and the main sequence is shown in Figure~\ref{fig:CEERS_MS}. As a cross-check, in this appendix we now also define the main sequence using CEERS data alone and verify that the key results of this paper based on the main sequence method remain unchanged, irrespective of whether we use the main sequence from CEERS or the one from \citet{Popesso2023}.

In order to define a main sequence using CEERS data alone, we construct a star-forming main sequence within each redshift bin using the full CEERS galaxy sample, including barred disks, unbarred disks, highly inclined systems, non-disks, etc. Galaxies are grouped into stellar mass bins of width 0.5 dex, requiring a minimum of 10 galaxies per bin. Within each mass bin, the median log SFR is computed and the resulting values are linearly interpolated between bin centers to provide an initial estimate of the main sequence.

\begin{figure*}[h]
    \centering
    \includegraphics[width=1\textwidth]{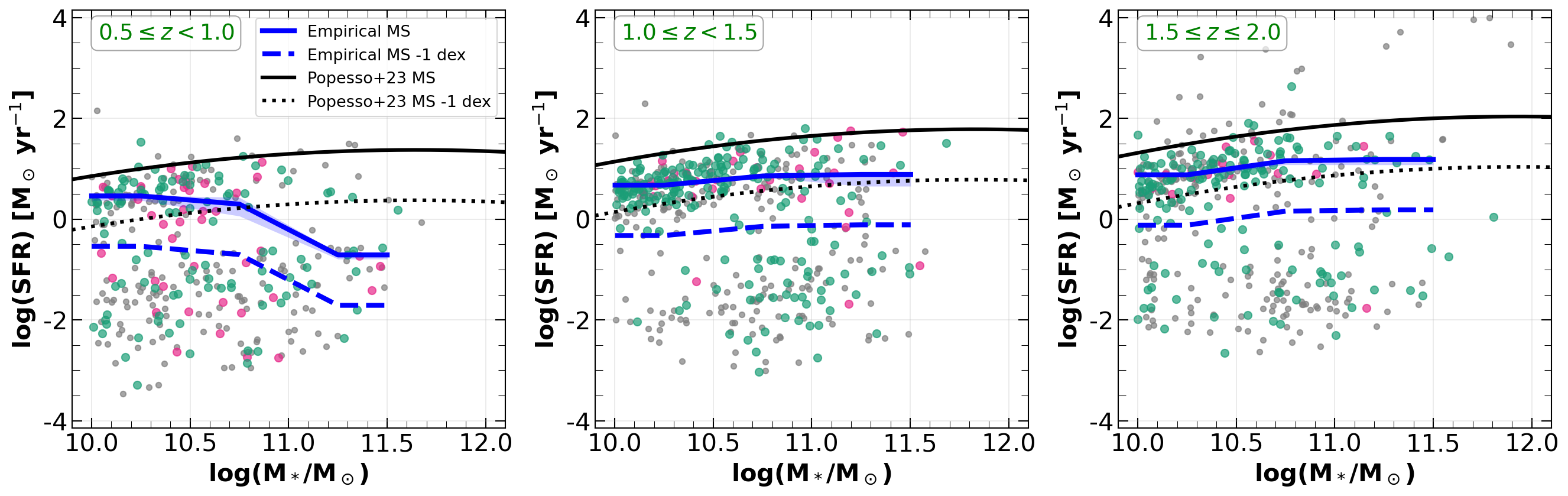}
    \caption{
    We define the main sequence using CEERS data alone. The solid blue line shows the main sequence derived using CEERS data, and the dashed blue line shows the CEERS main sequence -1 dex. Unbarred disk galaxies are shown in green, barred disk galaxies in magenta, and galaxies above our stellar mass cut that do not fall within those definitions in grey (e.g., spheroids, highly inclined disks, etc.). For comparison, the solid black line shows the main sequence published in \citet{Popesso2023} Equation 10, while the dotted black line shows the \citet{Popesso2023} main sequence -1 dex.}
    \label{fig:popesso_ms}
\end{figure*}

We then isolate the star-forming population by excluding galaxies lying more than 0.5 dex below the aforementioned main sequence. The main sequence is next recomputed using only the remaining galaxies, yielding the final CEERS main sequence. Uncertainties on the empirical main sequence are estimated via bootstrap resampling. In each stellar-mass bin, 1000 bootstrap realizations are generated by resampling galaxies with replacement, and the median log SFR is recalculated for each realization. The blue lines shown in Figure~\ref{fig:popesso_ms} correspond to the final CEERS main sequence, the shaded blue regions correspond to the 18th--64th percentile range of the bootstrapped median values, and the black lines show the parameterized main sequence from \citet{Popesso2023}. As shown in Figure~\ref{fig:popesso_ms}, the \citet{Popesso2023} main sequence displays a slight offset to our empirical relation, possibly tied to SFR calibrations.

\begin{figure}
\centering
\includegraphics[width=0.8\linewidth]{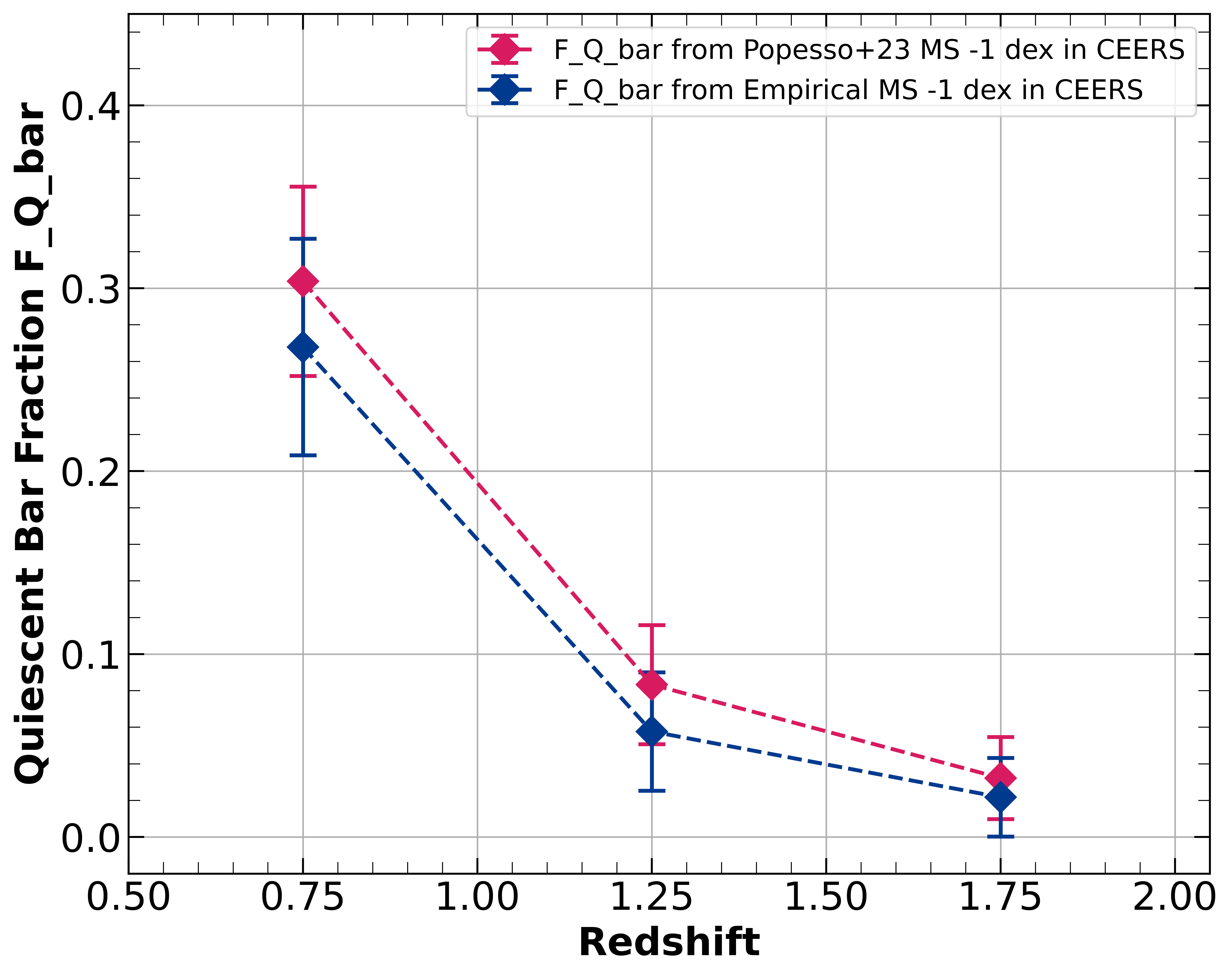}
\caption{The observed bar fraction among quiescent disk galaxies with $M_\star > 10^{10}\,M_\odot$ ($F_{\mathrm{Q\_bar}}$) is plotted from $z \sim 2$ to $z \sim 0$ based on CEERS data. The dashed blue line shows the $F_{\mathrm{Q\_bar}}$ derived using the CEERS main sequence to select quiescent galaxies and the dashed magenta line shows the $F_{\mathrm{Q\_bar}}$ derived using the \citet{Popesso2023} main sequence to select quiescent galaxies. The steep rise in $F_{\mathrm{Q\_bar}}$ remains unchanged, verifying that the key results of this paper based on the main sequence method remain unchanged, irrespective of whether we use the main sequence from CEERS or the one from \citet{Popesso2023}.}
\label{fig:popesso_frac}
\end{figure}

To assess the robustness of our results to the choice of main sequence, we repeat our analysis on the quiescent bar fraction ($F_{\mathrm{Q\_bar}}$) using the CEERS main sequence. As described in Section~\ref{sec:quiescent id}, we define quiescent galaxies relative to the main sequence. Galaxies with log SFR more than 1 dex below this main sequence are classified as quiescent. We compare the quiescent bar fractions derived using the CEERS main sequence (dashed blue line) and the published \citet{Popesso2023} main sequence (dashed magenta line) in Figure~\ref{fig:popesso_frac}. The quiescent bar fractions are in very good agreement, with both methods showing a steep rise in $F_{\mathrm{Q\_bar}}$ from $z\sim2$ to $z\sim0.5$, irrespective of whether we use the main sequence from CEERS or the one from \citet{Popesso2023}.

\newpage
\bibliography{sfr_bar}{}
\bibliographystyle{aasjournalv7.1}




\end{document}